\documentclass[aps,prb,reprint,groupedaddress]{revtex4-2}%
\usepackage{tabu}
\usepackage{graphicx}
\usepackage{epstopdf}
\usepackage{dcolumn}
\usepackage{bm}
\usepackage[breaklinks=true]{hyperref}
\usepackage{hyperref}
\usepackage{tabularx}
\usepackage{wrapfig}
\usepackage{stackengine}
\usepackage{float}
\usepackage{hyperref}
\usepackage{hhline}
\usepackage{bm}
\usepackage[table]{xcolor}
\usepackage{tabularx}
\usepackage{ltablex}
\usepackage{booktabs}
\usepackage{fancyhdr}
\usepackage{amsmath}
\usepackage{amssymb}
\usepackage{epsfig}
\usepackage{subfigure}
\usepackage{graphicx}
\usepackage[rightcaption]{sidecap}
\usepackage{float}
\usepackage{here}
\usepackage{upgreek}
\usepackage{dcolumn}
\usepackage{har2nat}
\usepackage{rotating}
\usepackage{multirow}
\usepackage{comment}
\usepackage{epigraph}
\usepackage[export]{adjustbox}
\usepackage[toc,page]{appendix}
\usepackage{enumitem}
\usepackage{sidecap}
\usepackage{pifont}
\usepackage{footnotebackref}
\usepackage[symbol]{footmisc}
\usepackage{mhchem}
\usepackage{color}
\usepackage{MnSymbol}
\usepackage{marginnote}
\usepackage{appendix}
\usepackage{float}
\usepackage{here}
\usepackage{longtable}
\usepackage{multirow}
\usepackage{placeins}
\usepackage{amsfonts}%
\providecommand{\U}[1]{\protect\rule{.1in}{.1in}}

\hypersetup{colorlinks=true, citecolor=blue, filecolor=blue, linkcolor=blue,urlcolor=black}
\sidecaptionvpos{figure}{c}
\definecolor{ao(english)}{rgb}{0.0, 0.5, 0.0}
\definecolor{alizarin}{rgb}{0.82, 0.1, 0.26}
\definecolor{black}{rgb}{0.0, 0.0, 0.0}
\begin{document}

\preprint{APS/123-QED}

\title{Electrically induced strong modulation of magnons transport \\in ultrathin magnetic insulator films}

\author{J. Liu}

\affiliation{Physics of Nanodevices, Zernike Institute for Advanced Materials, University of Groningen, Nijenborgh 4, 9747 AG Groningen, The Netherlands}
	
\author{X-Y. Wei}
\email{x.wei@rug.nl}
\affiliation{Physics of Nanodevices, Zernike Institute for Advanced Materials, University of Groningen, Nijenborgh 4, 9747 AG Groningen, The Netherlands}%
	
\author{G. E. W. Bauer}
\affiliation{Zernike Institute for Advanced Materials, University of Groningen, Nijenborgh 4, 9747 AG Groningen, The Netherlands}%
\affiliation{WPI-AIMR $\&$ Institute for Materials Research $\&$ CSRN, Tohoku University, Sendai 980-8577, Japan}%
	
\author{J. Ben Youssef}
\affiliation{LabSTICC, UMR CNRS 6285, Universit\'e de Bretagne Occidentale, 6 Avenue Le Gorgeu, 29238 Brest Cedex 3, France}%
\author{B. J. van Wees}
\email{b.j.van.wees@rug.nl}
\affiliation{Physics of Nanodevices, Zernike Institute for Advanced Materials, University of Groningen, Nijenborgh 4, 9747 AG Groningen, The Netherlands}

\date{\today}

\begin{abstract}
Magnon transport through a magnetic insulator can be controlled by
current-biased heavy-metal gates that modulate the magnon conductivity via the
magnon density. Here, we report nonlinear modulation effects in 10$\,$nm thick
yttrium iron garnet (YIG) films. The modulation efficiency is larger than
40\%/mA. The spin transport signal at high DC current density (2.2$\times
10^{11}\,$A/m$^{2}$) saturates for a 400$\,$nm wide Pt gate, which indicates
that even at high current levels a magnetic instability cannot be reached in
spite of the high magnetic quality of the films.

\end{abstract}
\maketitle

\section{Introduction}

\label{section:intro_3Terminal}

Magnons, i.e. the quanta of spin waves, are carriers of information with
properties that are attractive for applications \cite{magnon_spintronics}.
Magnons propagate in ferro-, ferri, antiferro-, and even paramagnetic electric
insulators without Joule heating \cite{ferro, ferri, antiferro, 1811.11972}.
The ferrimagnet yttrium iron garnet (YIG) is to date the best platform for
magnon spintronics due to its low Gilbert damping and high Curie temperature.
In YIG, magnons can be excited thermally and electrically and can cover long
distances \cite{spin_seebeck, tunable_spin_hall, Cornelissen2015}. An electric
current $I$ in a thin-film platinum contact generates a spin accumulation at
the Pt$|$YIG interface, which injects magnons into YIG. The latter diffuse
into the magnet and when reaching another Pt contact generate a voltage $V$ by
the inverse spin Hall effect. The non-local resistance $R_{\mathrm{nl}}=V/I$
can be modulated by a third Pt film, as demonstrated for a 210$\,$nm thick YIG
film \cite{PhysRevLett.120.097702}. This three terminal device is a magnon
transistor. The left and right ones inject and detect magnons thus form a
\emph{source} and a \emph{drain}, respectively. Sending a current though the
middle strip or \emph{gate} modulate the source-drain signal by the magnon
density in the transport channel.

Chumak et al. \cite{chumak_transistor} achieved magnon transistor action by
controlling the magnon scattering in a magnonic crystal by a magnetic field.
Our device operates by modulating the magnon conductivity of a YIG thin film
$\sigma_{\mathrm{m}}$ electrically. Similar to the Drude formula for
electrons, the magnon conductivity
\begin{equation}
\sigma_{\mathrm{m}}=\hbar\frac{n_{\mathrm{m}}\tau_{\mathrm{m}}}{m_{\mathrm{m}%
}},
\end{equation}
on the magnon density $n_{\mathrm{m}}$, where $\tau_{\mathrm{m}}$ is the
scattering time and $m_{\mathrm{m}}=\hbar^{2}/\left(  2J_{\mathrm{s}}\right)
$ is the effective mass that is governed by the spin wave stiffness
$J_{\mathrm{s}}$.

\begin{figure}[tb]
\vspace{0pt} \includegraphics[width=1\linewidth]{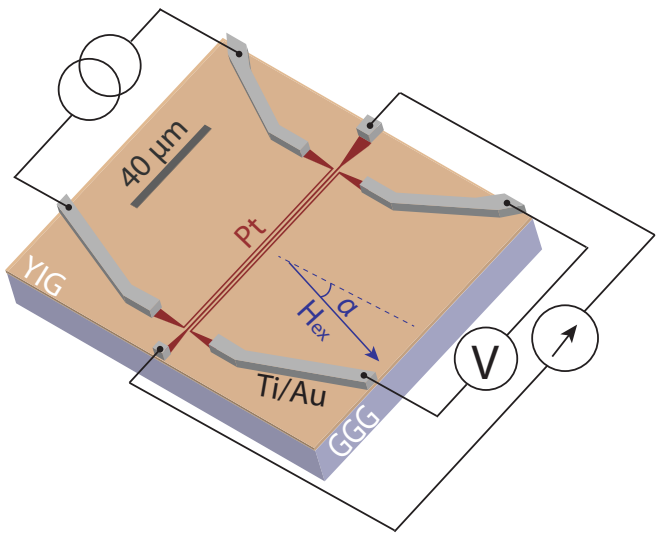}
\vspace{-10pt}\caption{Sample schematic\textbf{: }A 10$\,$nm YIG film grown
epitaxially on top of a GGG substrate. The sputtered Pt (red) strips with
thickness of 9$\,$nm are contacted by Ti/Au leads (grey). A low-frequency AC
current with rms value of $I_{\mathrm{ac}}$ in the left Pt strip injects
magnons. We measure both the first and second harmonic voltages over the right
Pt strip by a lock-in technique. The DC current through the middle Pt middle
gate modulated the source-drain signal. An external magnetic field
$H_{\mathrm{ex}}$ orients the in-plane YIG magnetization at an angle $\alpha$.
The dark-grey rectangle is a 40 $\,\mathrm{\mu}$m scale bar. Typically,
$\mu_{0}H_{\mathrm{ex}}=50\,$mT.}%
\label{fig:10nmDevice123Optical}%
\end{figure}

The present study is motivated by the wish to improve the modulation
efficiency of the previous device \cite{PhysRevLett.120.097702}. This can be
achieved simply by a thinner YIG\ film, since for the same number of injected
magnons, the magnon density in the source-drain transport channel should be
larger \cite{1812.01334}. To this end we grew an ultra-thin YIG film by
liquid-phase expitaxy with thickness of 10$\,$nm with great case in order to
not sacrifice the low Gilbert damping of the thicker film. The observed
modulation of the nonlocal signal reaches 200$\,\%$ corresponding to a
modulation efficiency per DC current unit exceed 40\%/mA, which is 20 times
larger than for the 210$\,$nm YIG \cite{PhysRevLett.120.097702}. A similar
enhancement has been reported for a $13\,$nm thick YIG film grown by pulsed
laser deposition and larger Gilbert damping \cite{1812.01334}. The authors
interpret an observed non-linearity in the gate-current dependence in terms of
a diverging magnon conductivity by a spin Hall current-induced antidamping of
the magnetization dynamics under the gate. Based on the observed dependence of
the modulation on the gate width and geometry we believe that the physics is
more complicated.

This paper is organized as follows: Section
\ref{section:ExperimenalDetails_3Terminal} addresses the device configuration,
fabrication details and measurement methods. In Section
\ref{section:Discussion_3Terminal}, we first compare the nonlocal signals in
10$\,$nm and 210$\,$nm thick YIG films. We then discuss the non-linearities
that in contrast to the previous report \cite{1812.01334} saturate, discuss
other device configurations, and show results of spin Hall magnetoresistance
measurements of the Pt$|$YIG interface at high gate currents. In Section
\ref{section:Discussion_3Terminal}, we compare our results with those reported
by Wimmer et al. \cite{1812.01334}.

\section{Experimental details}

\label{section:ExperimenalDetails_3Terminal}

The magnon transistors as depicted in Fig.$\,$\ref{fig:10nmDevice123Optical}
are fabricated on 10$\,$nm thick single crystal yttrium iron garnet (YIG)
films. The film is grown by liquid phase epitaxy (LPE) on top of a
500$\,\mathrm{\mu}$m thickness single crystal (110) gadolinium gallium garnet
(GGG, Gd$_{3}$Ga$_{5}$O$_{12}$) substrate at the Universit\'{e} de Bretagne
Occidentale in Brest, France. The saturation magnetization is $\mu
_{0}M_{\mathrm{s}}=174\pm4\,$mT. The Gilbert damping parameter of the in-plane
magnetized film is $\alpha_{G}=5.2\times10^{-4}$. All Pt strips, including the
magnon injector, modulator and detector, are sputtered with thickness of 9 nm,
patterned by electron beam lithography. Ti$|$Au layers with thicknesses of
5$|$75$\,$nm are deposited by e-beam evaporation. The center-to-center
distance between the injector and detector is 3$\,\mathrm{\mu}$m. The length
and width of the Pt strips for 3 measured devices are listed in Table
\ref{TableDevice123}, but we focus on Device 1. Results for a fourth device
with 7.9 nm thickness YIG are summarized in Appendix B. \textit{ }The sample
is positioned between a pair of magnetic poles and rotated by a step motor.
The magnetic field $\mathbf{H}_{\mathrm{ex}}$ orients the soft magnetization
$\mathbf{M}_{0}\Vert\mathbf{H}_{\mathrm{ex}}$ in the film plane at an angle
$\alpha$ with respect to the Pt strips as shown in Fig.$\,$%
\ref{fig:10nmDevice123Optical}.

A low-frequency AC current through the magnon injector with an rms-amplitude
of $I_{\mathrm{AC}}$, thereby injecting magnons electrically and thermally.
The resulting magnon spin currents are measured as the first and second
harmonic signals at the magnon detector with a lock-in technique,
respectively. A DC current $I_{\mathrm{DC}}$ is applied to the gate in order
to modulate the magnon spin conductivity and the corresponding nonlocal signals.

\begin{table}[tb]
\caption{Dimensions of the injector/modulator/detector Pt strips and selected
observations. The centers of injectors and detectors are separated by
3\thinspace$\mathrm{\mu m}$ and the Pt film thicknesses is 9 nm in all
samples}%
\label{TableDevice123}%
\vspace{3pt} \centering\setlength{\extrarowheight}{-2pt}
\begin{tabular}
[c]{cccc}%
\hhline{====}\vspace{0pt} &  &  & \\
Device & 1 & 2 & 3\\\hline
\vspace{0pt} &  &  & \\
Length ($\mu$m) & 80/84/80 & 20/24/20 & 20/24/20\\
\vspace{-0.5pt} &  &  & \\
Width ($\mu$m) & 0.4/0.4/0.4 & 0.4/0.8/0.4 & 0.4/1.2/0.4\\
\vspace{-0.5pt} &  &  & \\
$I_{\text{ac}}$ ($\mu$A) & 200 & 500 & 500\\
\vspace{-0.5pt} &  &  & \\
$I_{\text{dc}}$ ($m$A) & -1.5$\sim$1.5 & -2.0$\sim$2.0 & -2.25$\sim$2.25\\
` \vspace{-0.5pt} &  &  & \\
$R_{\text{nl}}^{1\omega}$ at $I_{\text{dc}}$=0 ($\Omega$/m) & 198 & 1044 &
160\\
\vspace{-0.5pt} &  &  & \\
Modulation &  &  & \\
efficiency (\%/mA) & 40.4 & 87 & 75\\
\vspace{0pt} &  &  & \\
\hhline{====} &  &  &
\end{tabular}
\end{table}

\begin{figure}[tb]
\includegraphics[width=1\linewidth]{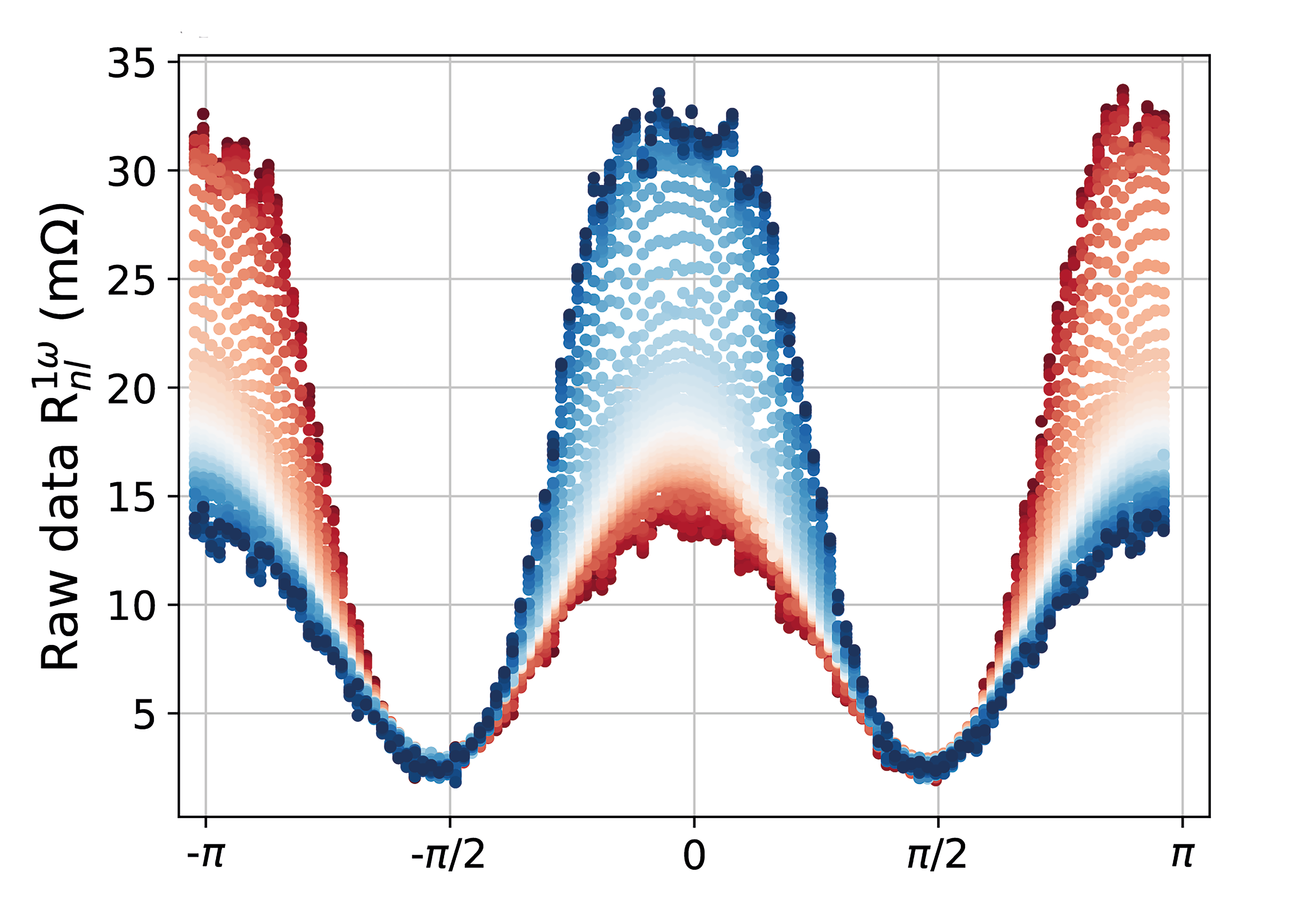} \caption{Angle dependent
$R_{\mathrm{nl}}^{1\omega}$\textbf{.} Raw data of the first harmonic signals
$R_{\mathrm{nl}}^{1\omega}$ at different DC gate currents with offset from
inductive/capacitive coupling (at $\alpha=\pm\pi/2$). The color gradient from
red to blue represents DC currents from -1500$\,\mathrm{\mu}$A to
+1500$\,\mathrm{\mu}$A with a step size of 50$\,\mathrm{\mu}$A.}%
\label{fig:10nmDevice1V1w}%
\end{figure}

The observed angle-dependent first harmonic signals of Device 1 are shown in
Fig.$\,$\ref{fig:10nmDevice1V1w}: Colors, from red to blue code the nonlocal
signals recorded for $I_{\mathrm{DC}}$ from -1500$\,\mathrm{\mu}$A to
+1500$\,\mathrm{\mu}$A. The white dataset in the center for $I_{\mathrm{DC}%
}=0$ has a typical $\cos^{2}\alpha$ dependence, i.e. the product of injection
and detection efficiencies \cite{Cornelissen2015}. The DC bias modulates the
magnitude and the angle dependence much more prominently than for a 210$\,$nm
thick YIG film \cite{PhysRevLett.120.097702}, especially at the largest
currents of -/+1500$\,\mathrm{\mu}$A (the darkest red/blue) and $\alpha
\approx0$ and $\alpha\approx\pm\pi$. The gate annihilates magnons in YIG when
the spin accumulation is parallel to the magnetic field but creates them when
antiparallel, suppression and enhancing $R_{\text{\textrm{nl}}}^{1\omega}$ is
suppressed, respectively. The DC current enhances the signal by more than a
factor of 2. Also the second harmonic signals are strongly modulated by the
gate current (not shown), but more difficult to interpret since depending not
only on the magnon density but also on the temperature profiles in the magnet.
We therefore do not discuss them here.

\section{Results and Discussion}

\label{section:Discussion_3Terminal}

\subsection{Dependence of the nonlocal signals on YIG film thickness}

The nonlocal signals for 10$\,$nm (Device 1 in Table \ref{TableDevice123}) and
210$\,$nm thick YIG films with the same injector-to-detector distance
(3$\,\mathrm{\mu}$m) are compared in Table \ref{Table210nm10nmNonlocal}. The
nonlocal resistances scale with the length of the Pt strips. The ultra-thin
gated but unbiased 10$\,$nm YIG sample shows a larger non-local signal than
the thick one without gate, even though a passive central gate is a spin sink.
This result is consistent with the thickness-dependence reported for films
from 100$\,$nm up to 50$\,\mathrm{\mu}$m $\,$\cite{PhysRevB.94.174437}, but
counterintuitive since a thinner film should have a higher impedance. It
cannot be explained by either the magnon chemical potential model
\cite{PhysRevB.94.014412} nor viscous magnon flow \cite{1903.02790}. On the
other hand, the second harmonic spin Seebeck signal in 10$\,$nm thick YIG (not
shown) is much smaller in the 10 nm than in the 210$\,$nm film. The thickness
dependence of the nonlocal magnon transport remains unexplained. We may
speculate for example about a the existence of highly efficient surface
transport channels that dominate in ultra thin films. The thickness dependence
of the nonlocal signal will be discussed in a future paper with more details.

\begin{table}[h]
\caption{Comparison of the first-harmonic nonlocal signals in 10$\,$nm and
210$\,$nm thick YIG films.}%
\label{Table210nm10nmNonlocal}
\vspace{3pt} \centering
\setlength{\extrarowheight}{0pt}
\begin{tabular}
[c]{cccc}%
\hhline{====}\vspace{-5pt} &  &  & \\
YIG thickness (nm) & 10 & 210{\cite{Cornelissen2015}} & \\\hline
\vspace{-2pt} &  &  & \\
$R^{1\omega}_{\text{nl}}$ ($\Omega$m$^{-1}$) & 198 & 140 & \\
\vspace{-5pt} &  &  & \\
$R^{2\omega}_{\text{nl}}$ (MVA$^{-2}$m$^{-1}$) & 0.09 & 1.35 & \\
\vspace{-2pt} &  &  & \\
\hhline{====} &  &  &
\end{tabular}
\end{table}

\subsection{Saturation in the injector/modulator/detector geometry for a
400$\,$nm wide gate}

The nonlocal resistances $R_{\mathrm{nl}}^{\mathrm{1}\omega}$ are
trigonometric functions of the magnetic field angle $\alpha$ that reflect the
electrical magnon injection and detection efficiencies \cite{Cornelissen2015}.
The angle-dependent first-harmonic nonlocal resistances are well described by
\begin{equation}
R_{\mathrm{nl}}^{\mathrm{1}\omega}(\alpha)={C_{1}}\sigma_{\mathrm{m}}%
^{1\omega}(\alpha)\cos^{2}\alpha,
\end{equation}
where $C_{1}$ is a charge-spin conversion efficiency parameter of the electric
spin injection and detection. In the limit of weak excitation, the magnon spin
conductivity depends linearly on the magnon density which is again
proportional to the and injection current. We also include a quadratic term
that does not depend on the current direction and is caused by Joule heating.
Hence
\begin{equation}
\sigma_{\mathrm{m}}^{1\omega}(\alpha)=\sigma_{\mathrm{m}}^{0}+\Delta
\sigma_{\mathrm{SHE}}I_{\mathrm{DC}}\cos\alpha+\Delta\sigma_{\mathrm{J}%
}I_{\mathrm{DC}}^{2}, \label{eq_MagnonSpinConductivity}%
\end{equation}
where $I_{\mathrm{DC}}$ is the DC current in the modulator, $\sigma
_{\mathrm{m}}^{0}$ is the magnon spin conductivity at thermal equilibrium,
$\Delta\sigma_{\mathrm{J}}$ and $\Delta\sigma_{\mathrm{SHE}}$ are parameters
that can be fitted to the observations.

\begin{figure*}[t]
\centering
\begin{subfigure}
		\centering
		\includegraphics[width=0.38\linewidth]{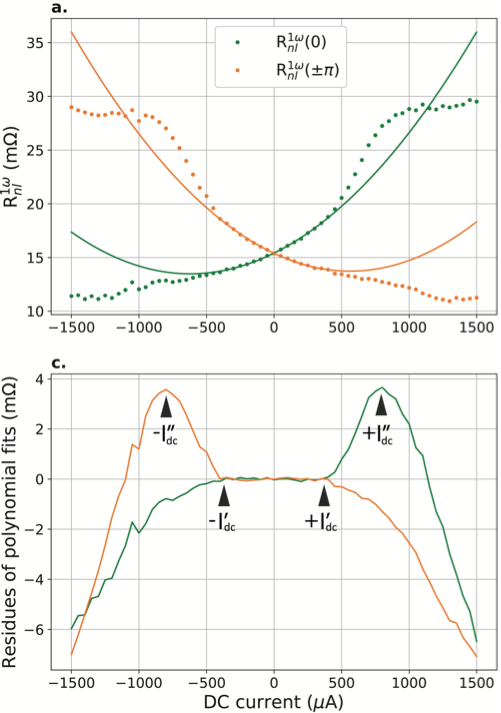}
	\end{subfigure}\hspace*{0.9em} \begin{subfigure}
		\centering
		\includegraphics[width=0.38\linewidth]{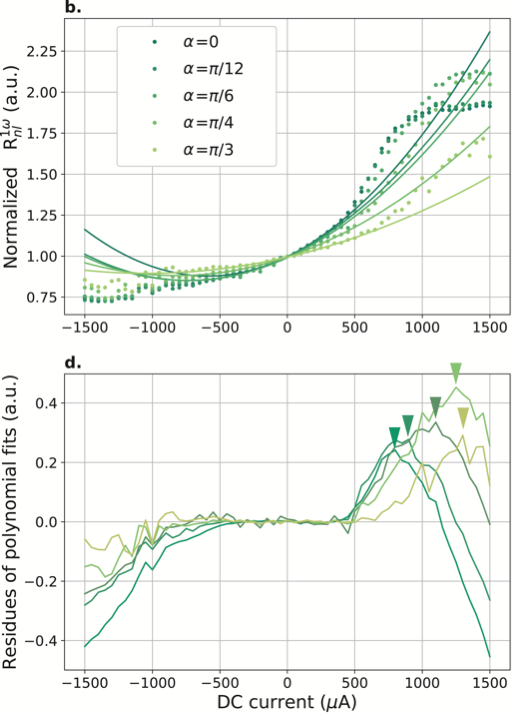}
	\end{subfigure}
\caption{Analysis of $R_{\mathrm{nl}}^{1\omega}$ at specific angles $\alpha$
for Device 1. The dots are the experimental data and the lines with the same
color are quadratic fits for small currents. a.
\textcolor{ao(english)} {$\bullet$} $R_{\mathrm{nl}}^{1\omega}(0)$) and
$\alpha+\phi^{1\omega}=\pm\pi$ ( \textcolor{orange}{$\bullet$} $R_{\mathrm{nl}%
}^{1\omega}(\pm\pi)$) as a function of DC gate current $I_{\mathrm{DC}}$. b.
Data for $\alpha=\left\{  0,\pi/12,\pi/6,\pi/4,\pi/3\right\}  $ normalized by
the amplitudes at $I_{\mathrm{DC}}=0$ (green dots with different brightness).
The deviations from the fits in a. and b. are plotted in c. and d.,
respectively. }%
\label{fig:10nmDevice1V1wMunich}%
\end{figure*}

We extract the non-local resistances at specific angles from Fig.$\,$%
\ref{fig:10nmDevice1V1w} as a function of $I_{\mathrm{DC}}$, subtracting a
constant offset at $\alpha=\pm\pi/2$ from the measured $R_{\mathrm{nl}%
}^{\mathrm{1}\omega}(\alpha)$ that is caused by inductive/capacitive coupling.
The signals at the angles $\alpha=0,\pm\pi$ are shown in Figs.$\,$%
\ref{fig:10nmDevice1V1wMunich}a as well as normalized ones for $\alpha
=0,\frac{\pi}{12}$, $\frac{\pi}{6}$, $\frac{\pi}{4}$ and $\frac{\pi}{3}$ in
Figs.$\,$\ref{fig:10nmDevice1V1wMunich}b. When $\left\vert I_{\mathrm{DC}%
}\right\vert <I_{\mathrm{DC}}^{\prime}=400\,\mathrm{\mu}$A (The current
density is $1.1\times10^{11}\,$A/m$^{2}$.), $R_{\mathrm{nl}}^{\mathrm{1}%
\omega}(\alpha)$ in Figs.$\,$\ref{fig:10nmDevice1V1wMunich}a is to a good
approximation a parabolic function of $I_{\mathrm{DC}}:$
\begin{equation}
R_{\mathrm{nl}}^{\mathrm{1}\omega}(I_{\mathrm{DC}})=\mathcal{P}_{0}^{1\omega
}+\mathcal{P}_{1}^{1\omega}I_{\mathrm{DC}}+\mathcal{P}_{2}^{1\omega
}I_{\mathrm{DC}}^{2}, \label{Eq_DeltaR_Idc}%
\end{equation}
with $\mathcal{P}_{1}^{1\omega}\sim6\,\Omega/$A and $\mathcal{P}_{2}^{1\omega
}$ $\sim5\times10^{3}\,\Omega/$A$^{2}$ ($\mathcal{P}_{0}^{1\omega}=1$ for the
normalized data in Figs.$\,$\ref{fig:10nmDevice1V1wMunich}b). The differences
between the observations and the fits of Figs.$\,$%
\ref{fig:10nmDevice1V1wMunich}a/b are given in Figs.$\,$%
\ref{fig:10nmDevice1V1wMunich}c/d , respectively. The data deviate from the
fits at the first threshold current $\left\vert I_{\mathrm{DC}}\right\vert
\gtrsim I_{\mathrm{DC}}^{\prime}=400\,\mathrm{\mu}$A (current density
$1.1\times10^{11}\,$A/m$^{2}$). At a second threshold $I_{\mathrm{DC}}%
^{\prime\prime}=800\,\mathrm{\mu}$A (current density $2.2\times10^{11}%
\,$A/m$^{2}$) the deviations from the polynomial fits show a \emph{maximum}
that we call an \emph{anomaly} for convenience. For $I_{\mathrm{DC}}<0$ the
parabolic model Eq.$\,$(\ref{eq_MagnonSpinConductivity}) predicts an increase
of the magnon conductivity by the parabolic term that models the magnon
injection by Joule heating. However, $R_{\mathrm{nl}}^{\mathrm{1}\omega}(0)$
deviates from this prediction for $I_{\mathrm{DC}}\lesssim-I_{\mathrm{DC}%
}^{\prime},$ i.e. at the same current level as for positive gate currents. The
experiments confirm that reversing the magnetic field is equivalent to
reversing the current direction. The threshold $I_{\mathrm{DC}}^{\prime\prime
}$ in Fig.$\,$\ref{fig:10nmDevice1V1wMunich}d increases with the angle
$\alpha,$ while $I_{\mathrm{DC}}^{\prime}$ remains constant. Since with
increasing $\alpha$ a higher current is required to inject the same number of
magnons, $I_{\mathrm{DC}}^{\prime\prime}$ appears to be related with the
magnon injection process, while $I_{\mathrm{DC}}^{\prime}$ is not.

\subsection{External field dependence of the saturation}

\begin{figure*}[tb]
\centering
\begin{subfigure}
		\centering
		\includegraphics[width=0.3\linewidth]{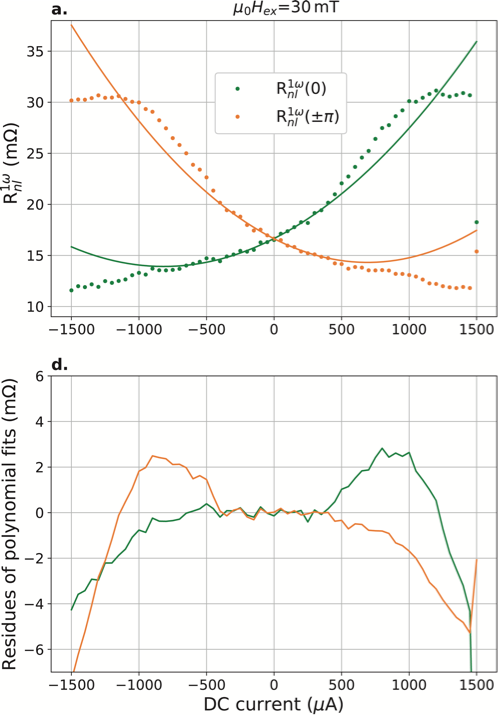}
	\end{subfigure}\hspace*{0.9em} \begin{subfigure}
		\centering
		\includegraphics[width=0.3\linewidth]{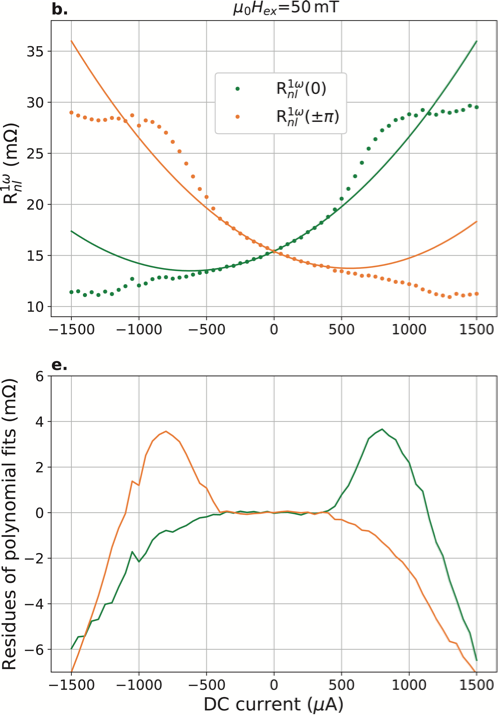}
	\end{subfigure}\hspace*{0.9em} \begin{subfigure}
		\centering
		\includegraphics[width=0.305\linewidth]{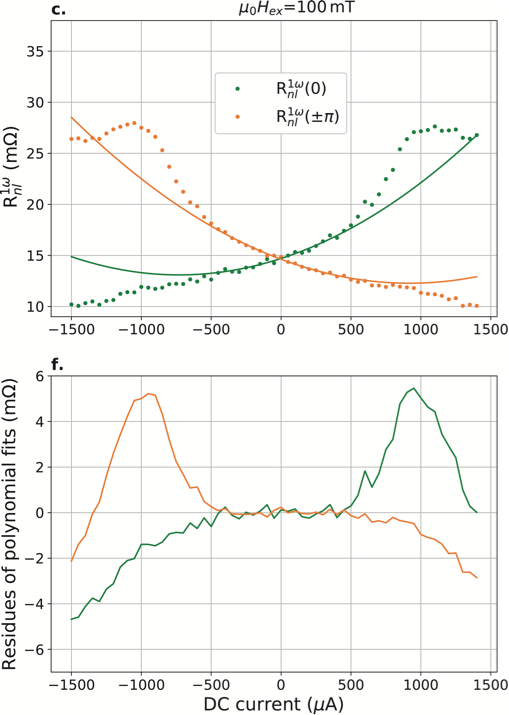}
	\end{subfigure}\hspace*{0.9em} \caption{Analysis of the non-parabolicties in
$R_{\text{\textrm{nl}}}^{1\omega}$ of Device 1 for different magnetic field.
We plot the deviations from the small-field parabolic fits for $\alpha=0$ (
\textcolor{ao(english)} {$\bullet$}$R_{\text{\textrm{nl}}}^{1\omega}(0)$) and
$\alpha=\pm\pi$ ( \textcolor{orange}{$\bullet$}$R_{\text{\textrm{nl}}%
}^{1\omega}(\pm\pi)$) as a function of the modulator current $I_{\mathrm{DC}}$
at 30$\,$mT, b 50$\,$mT and c 100$\,$mT.}%
\label{fig:10nmDevice1V1wMunichField}%
\end{figure*}

A field-dependent study can shed light on the possible effect of the magnon
gap or Kittel frequency
\begin{equation}
\omega_{k=0}=\gamma\sqrt{B_{0}\left(  B_{0}+\mu_{0}M_{\mathrm{s}}\right)  },
\label{Eq_magnondisperion}%
\end{equation}
on the anomaly $I_{\mathrm{DC}}^{\prime\prime}$. The results in Fig.$\,$%
\ref{fig:10nmDevice1V1wMunichField} show a slightly increased $I_{\mathrm{DC}%
}^{\prime\prime}$ with field from 700$\,\mathrm{\mu}$A to 1$\,$mA. This could
reflect a gap-induced reduction of the magnon number and conductivity.
$I_{\mathrm{DC}}^{\prime},$ which we found to not depend on the magnetization
angle above, remains also resilient against the magnetic field strength, however.

\subsection{Spin Hall magnetoresistance of the 400$\,$nm wide Pt strip}

\begin{figure}[tb]
\includegraphics[width=1\linewidth]{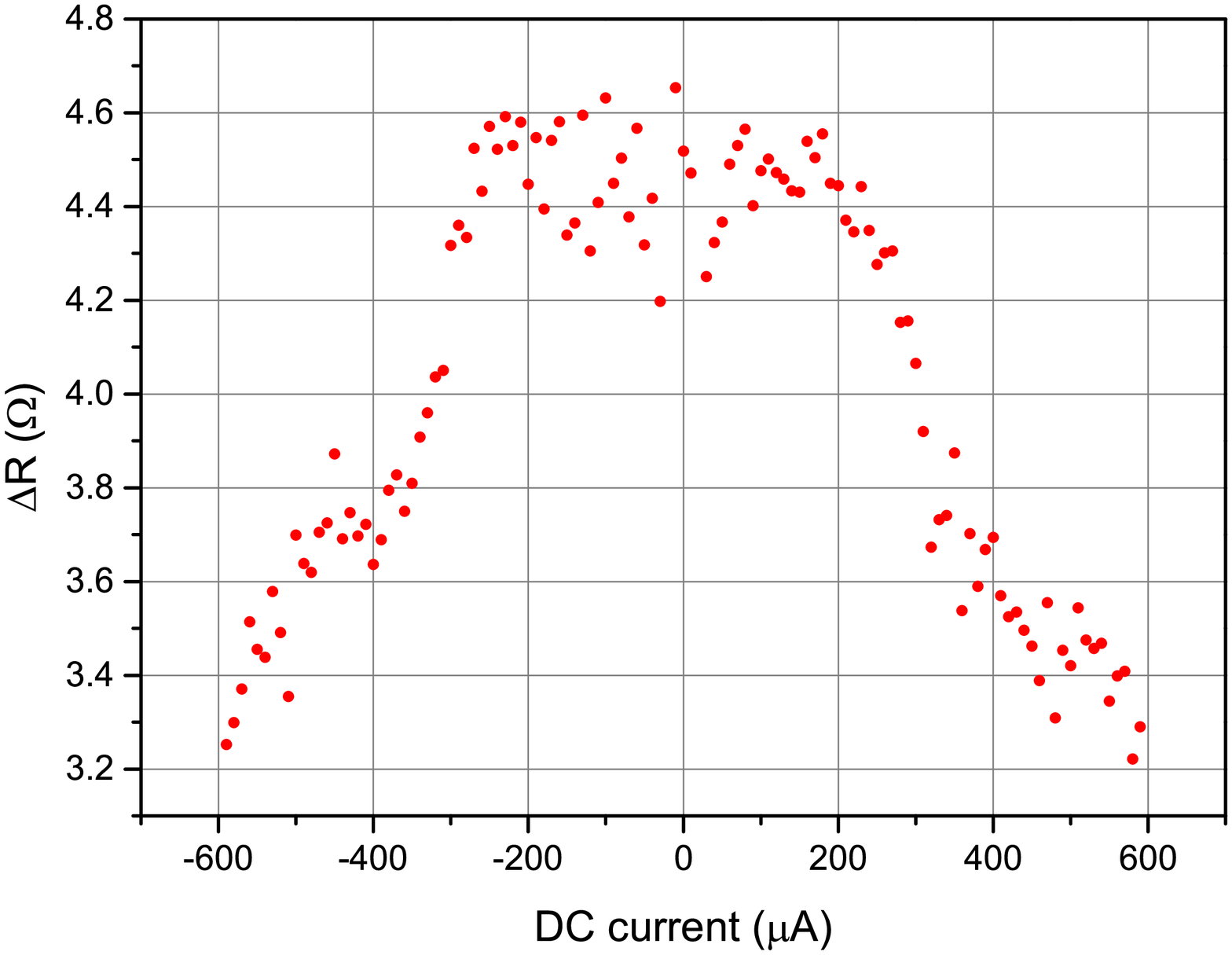} \caption{(Longitudinal) spin Hall
magnetoresistance of the 400 nm wide central Pt gate at DC current from
$-590\,\mathrm{\mu}$A to $590\,\mathrm{\mu}$A.}%
\label{fig:SMR}%
\end{figure}

Next, we analyze the spin Hall magnetoresistance (SMR) of the 400 nm wide Pt
center gate \cite{SMR_DC} for an AC current of 20 $\mathrm{\mu}$A and a DC
current range from -590 $\mathrm{\mu}$A to 590 $\mathrm{\mu}$A. Fig.$\,$%
\ref{fig:SMR} shows that the resistance change $\Delta R$ decreases with DC
current $I_{\mathrm{DC}}$ with a threshold around $\pm400$ $\mathrm{\mu}$A,
close to $I_{\mathrm{DC}}^{\prime}$ introduced above.\textbf{ }Since the SMR
decreases with temperature \cite{SMR_tem} and appears to be correlated with
$I_{\mathrm{DC}}^{\prime},$ the first threshold in the non-local signal could
be heat-induced, consistent with its independence on the magnetic field
reported above.

\subsection{Exchanged source and gate contacts}

\begin{figure}[h]
\includegraphics[width=1\linewidth]{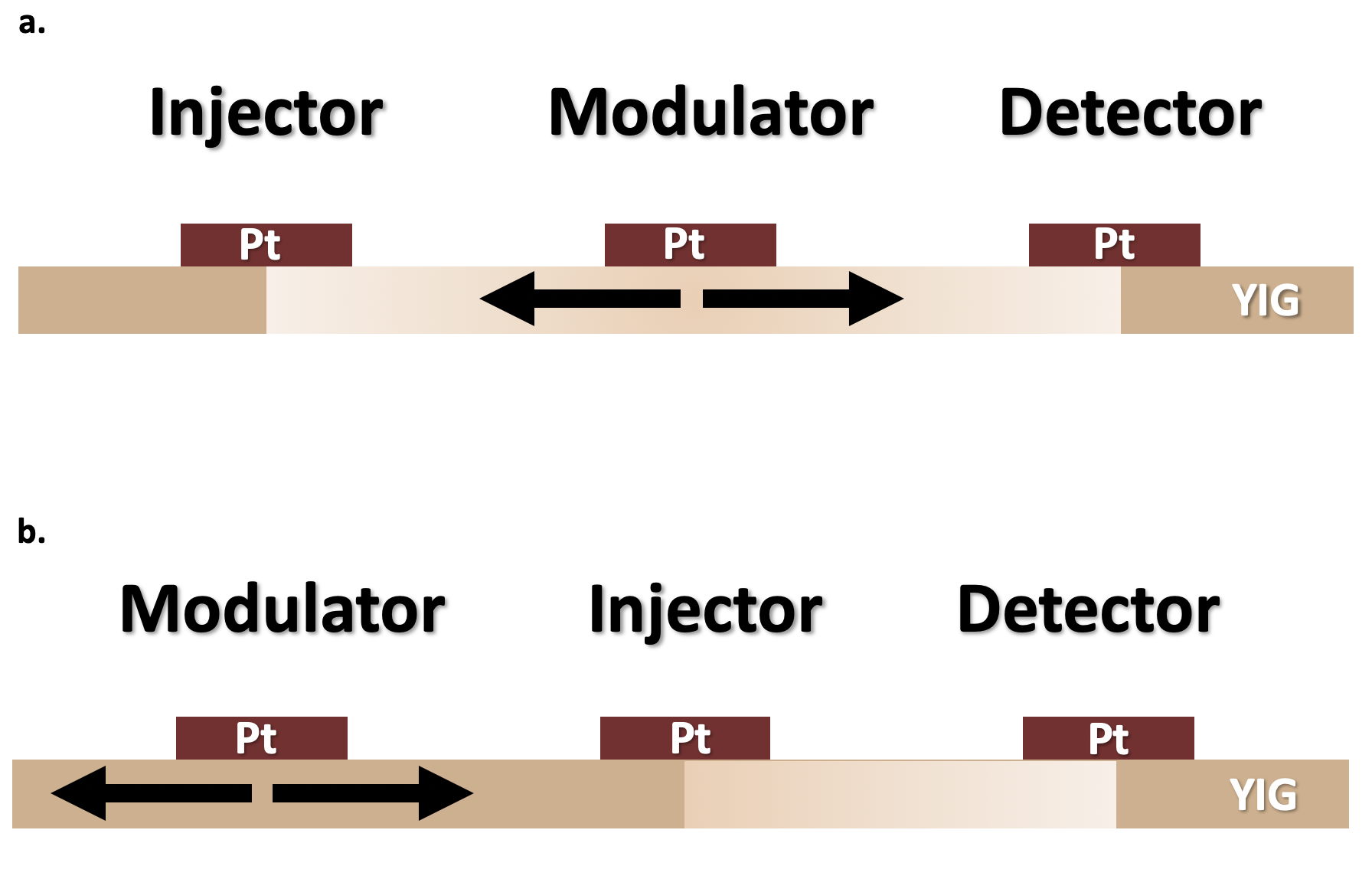} \caption{The
position of the modulating gate in two different configurations. The black
arrows represent the diffusion current of the magnons injected by the
modulator, while the lighter shaded regions indicate the source-drain path.
a. injector/modulator/detector configuration and b.
modulator/injector/detector configuration.}%
\label{fig:chemical_side}%
\end{figure}

\begin{figure}[h]
\includegraphics[width=1\linewidth]{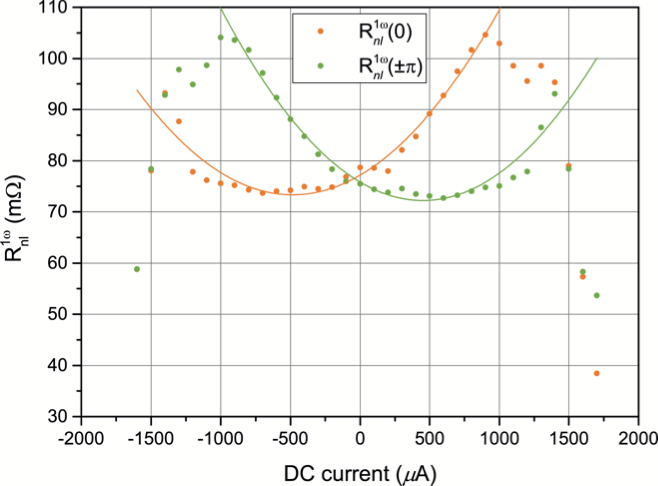} \caption{Analysis of
$R_{\mathrm{nl}}^{\mathrm{1}\omega}$ at specific angles for
modulator/injector/detector configuration. Relative amplitudes of the first
harmonic nonlocal signals of device 2 with 800 nm width modulator at
$\alpha=0$ ( \textcolor{orange} {$\bullet$}$R_{\mathrm{nl}}^{\mathrm{1}\omega
}(0)$) and $\alpha=\pm\pi$ (\textcolor{ao(english)}{$\bullet$}$R_{\mathrm{nl}%
}^{\mathrm{1}\omega}(\pm\pi)$) as a function of dc currents.}%
\label{fig:400nm_side}%
\end{figure}

In order collect more information on the anomalies observed in Figs.$\,$%
\ref{fig:10nmDevice1V1wMunich} we exchange roles of the Pt contacts from\ an
injector/modulator/detector to a modulator/injector/detector geometry in
Device 1 as sketched in Figs.$\,$\ref{fig:chemical_side}\textbf{b} and
specified in Table \ref{TableDevice5}. In this configuration the source-drain
current is not directly affected by an antidamping torque of the modulator.
The signal is larger because the injector and detector are now closer to each
other. The first harmonic signal for the new configuration in Fig.$\,$%
\ref{fig:400nm_side} is well represented by a parabola with $\mathcal{P}%
_{1}^{1\omega}\sim1.6\times10^{-2}\,\Omega/$A and $\mathcal{P}_{2}^{1\omega
}\sim18\,\Omega/$A$^{2}$ in Eq.$\,$(\ref{Eq_DeltaR_Idc}) for $\left\vert
I_{\mathrm{DC}}\right\vert <I_{\mathrm{DC}}^{\prime}=900\,\mathrm{\mu}$A.
$R_{\mathrm{nl}}^{\mathrm{1}\omega}(0)$ ($R_{\mathrm{nl}}^{\mathrm{1}\omega
}(\pm\pi)$) start to decreases for at currents $I_{\mathrm{DC}}%
=900\,\mathrm{\mu}$A (1400$\,\mathrm{\mu}$A) and $I_{\mathrm{DC}%
}=-1400\,\mathrm{\mu}$A (-900$\,\mathrm{\mu}$A). In contrast to the discussion
above, the deviations from the parabolic fit at $I_{\mathrm{DC}}^{\prime}$ are
negative so we cannot identify a $I_{\mathrm{DC}}^{\prime\prime}$.

In the modulator/injector/detector geometry, magnons injected by the modulator
first have to diffuse to the region between the injector and detector in order
to affect the magnon conductivity. The magnon chemical potential
$\mu_{\mathrm{m}}$ is a direct measure of the non-equilibrium magnon density
that obeys the spin diffusion equation $d^{2}\mu_{\mathrm{m}}/dx^{2}%
=\mu_{\mathrm{m}}/\lambda_{\mathrm{m}}^{2}$ \cite{Cornelissen2015} with a
magnon diffusion length of $\lambda_{\mathrm{m}}\sim10\,\mathrm{\mu m}$ at
room temperature.\ The magnon density in the source-drain channel amounts of
the side-modulator geometry should reduced to about 80\% of the value for the
center gate configuration. A larger current must be therefore be applied to
achieve the same magnon density. The additional heating may explain the
reduced performance.

\begin{table}[tb]
\caption{Geometry of modulator/injector/detector Pt strips and measurement
parameters.}%
\label{TableDevice5}%
\vspace{3pt} \centering
\setlength{\extrarowheight}{-2pt}
\begin{tabular}
[c]{ccccc}%
\hhline{=====} \vspace{0pt} &  &  &  & \\
Length ($\mu$m) & 20/24/20 &  &  & \\
\vspace{-0.5pt} &  &  &  & \\
Width ($\mu$m) & 0.4/0.8/0.4 &  &  & \\
\vspace{-0.5pt} &  &  &  & \\
Pt thickness (nm) & 9 &  &  & \\
\vspace{-0.5pt} &  &  &  & \\
$I_{\text{ac}}$ ($\mu$A) & 500 &  &  & \\
\vspace{-0.5pt} &  &  &  & \\
$I_{\text{dc}}$ ($m$A) & -1.6$\sim$1.7 &  &  & \\
\vspace{-0.5pt} &  &  &  & \\
Modulation efficiency (\%/$m$A) & 23.5 &  &  & \\
\vspace{0pt} &  &  &  & \\
\hhline{=====} &  &  &  &
\end{tabular}
\end{table}

\subsection{Modulator gate width dependence}

\begin{figure}[h]
\includegraphics[width=1\linewidth]{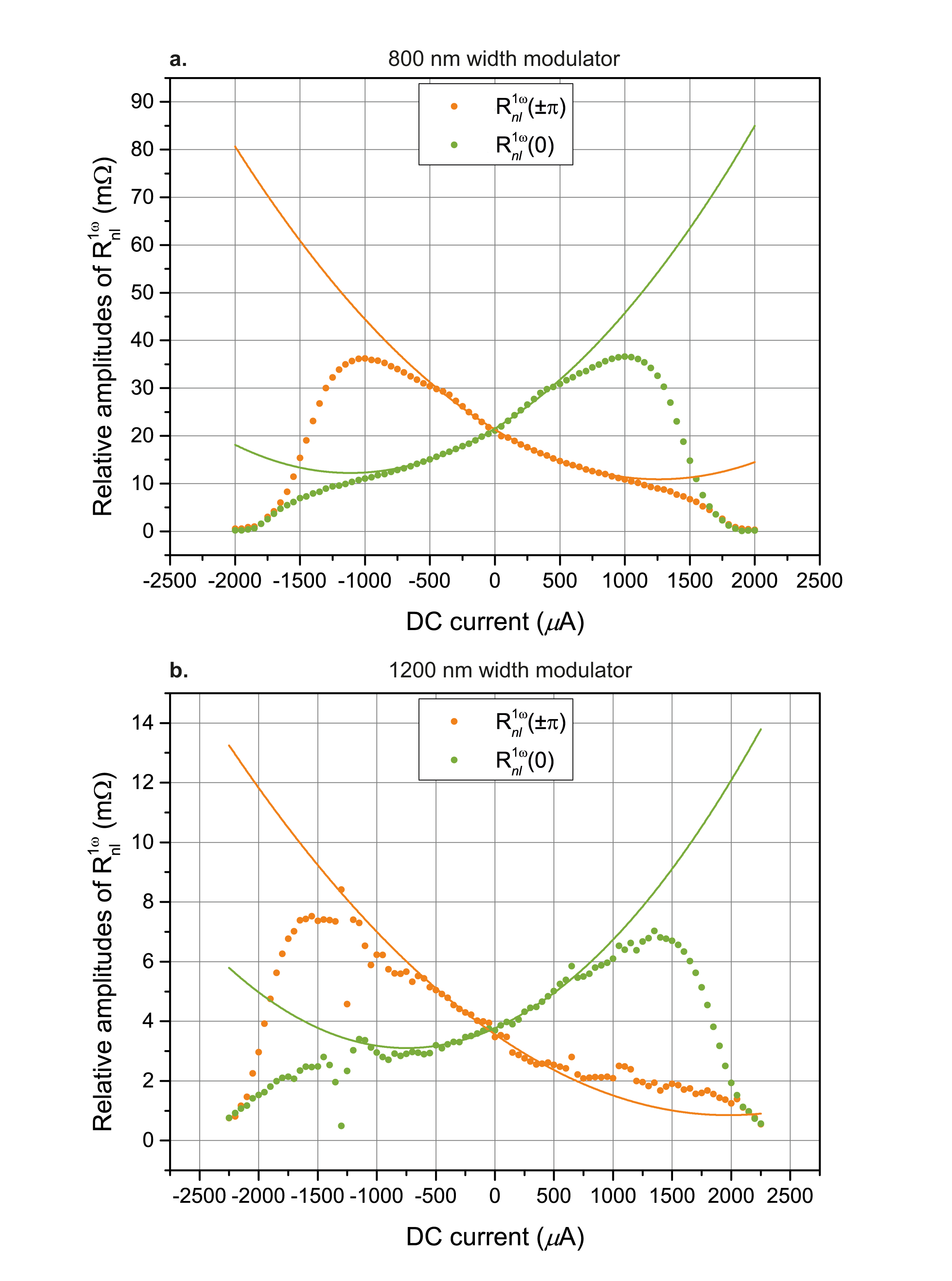} \caption{$R_{\mathrm{nl}%
}^{\mathrm{1}\omega}$ for Devices 2 and 3 with wider modulator gates as a
function of gate current $I_{\mathrm{DC}}$. a. The signals of Device 2 with
800 nm wide modulator at $\alpha=0$ ( \textcolor{ao(english)} {$\bullet$}
$R_{\mathrm{nl}}^{\mathrm{1}\omega}(0)$) and $\alpha=\pm\pi$ (
\textcolor{orange}{$\bullet$} $R_{\mathrm{nl}}^{\mathrm{1}\omega}(\pm\pi)$) as
a function of dc currents. b. As a function of gate current $I_{\mathrm{DC}}$,
but for Device 3 with 1200 nm wide modulator.}%
\label{fig:wider_modulator}%
\end{figure}

In Device 2 (800\thinspace nm wide modulator) and Device 3 (1200\thinspace nm
wide modulator) from Table I (see Fig.$\,$\ref{fig:wider_modulator}), a
saturation as in Device 1 (Fig.$\,$\ref{fig:10nmDevice1V1wMunich}a) is not
observed. The characteristics are similar to that of Device 1 in the
modulator/injector/detector configuration: $R_{\mathrm{nl}}^{\mathrm{1}\omega
}$ deviates from the simple magnon conductivity model at lower currents.
Device 2 deviates at currents (current densities) of 500\thinspace$\mathrm{\mu
A}$ ($0.7\times10^{11}\,$A/m$^{2}$) and starts to decrease at 1\thinspace mA
($1.4\times10^{11}\,$A/m$^{2}$). Device 3 deviates from $0.6\times10^{11}%
\,$A/m$^{2}$ and starts to decrease at $1.2\times10^{11}\,$A/m$^{2}$. They are
much lower than the current density corresponding to the saturation in Device
1 ($2.2\times10^{11}\,$A/m$^{2}$). How the width of the gate affects the
nonlinear effect also needs further investigation.

We also observe signal changes induced by a high DC current bias on Device 3
(see Appendix A) that indicate a transient change of the magnetic order of the
YIG film that may also cause the asymmetry between the data in Fig.$\,$%
\ref{fig:wider_modulator} for flipped current and magnetization directions.
Rather than blowing the sample up, we observed a strong increase of the
non-local signals. Since we have not been able to explain or repeat these
results we do not discuss them in the main text.

\section{Discussion and conclusions}

\label{section:Conclusion_3Terminal}We report large modulations of nonlocal
magnon transport in a 10$\,$nm thick YIG film by a DC current through a Pt
gate. For the injector/modulator/detector geometry, a threshold current
$I_{\mathrm{DC}}^{\prime}$ separates the low and high DC current regimes. The
enhancement of the magnon transport at currents $I>I_{\mathrm{DC}}^{\prime}$
indicates interesting physics such as current-induced self-oscillations of the
magnetic order. However, instead of a divergence that could indicate magnon
superfluidity, we observe a plateau at high current levels Figs.$\,$%
\ref{fig:10nmDevice1V1wMunich}a.

The differences between the data and a parabolic fit at low injection currents
sheds some light on what is happening. We clearly observe non-parabolicities
for both positive and negative currents, i.e. for both magnon injection and
extraction. At $I>I_{\mathrm{DC}}^{\prime}$ the signal is enhanced, i.e.
increases above the parabolic fit. This threshold is not sensitive to applied
magnetic fields and angles, which indicates a thermal (spin Seebeck) mechanism
for the enhancement of the conductivity as reported by C. Safranski et al.
\cite{spin_auto}\textit{. }The SMR data are suppressed around $I_{\mathrm{DC}%
}^{\prime},$ thereby supporting the hypthesis that Joule heating affects the
spin-transport at the interface. The residue of the polynomial fit in
Figs.$\,$\ref{fig:10nmDevice1V1wMunich}c shows a maximum, i.e. a peak at the
threshold current $I_{\mathrm{DC}}^{\prime\prime}$ (in one current direction),
and then decreases again. We cannot pinpoint the process that suppresses the
magnon conduction at high current levels to a certain mechanism, but it
appears to be spin-dependent since in contrast to $I_{\mathrm{DC}}^{\prime}$,
$I_{\mathrm{DC}}^{\prime\prime}$ depends strongly on the magnetic field
strength and direction.

When the modulator is in the center, the magnon transmission is affected by
thermal \cite{Amplification_of_Spin_Waves,Thermally_driven_spin_torques} or
electric \cite{PhysRevLett.113.197203,cite-key} spin-orbit torques as well as
spin absorption by the Pt gate. The situation is simplified for the
modulator/injector/detector geometry in so far that the modulator is only a
source of additional magnons that increase the injector-detector conductance.
$I_{\mathrm{DC}}^{\prime}$ is larger for this configuration, presumably
because the higher current level is required to generate the same density in
the source-drain channel by magnon diffusion. However, in contrast to the
center-gate configuration, the signal always stays under the parabolic fit.
This indicates that the magnon density is not the only parameter relevant for
magnon transport, confirming that a spin Seebeck torque from the Pt interface
plays an essential role.

Naively, we expected that for equal current densities the results should not
depend on the width of the gate. Nevertheless we find that widening the
central gates only decreases the signals relative to the polynomial fit. A
proper explanation of this result requires more research.

Summarizing, we observe a threshold behavior at currents $I>I_{\mathrm{DC}%
}^{\prime}$ that indicates that the film under the gate approaches an
instability, confirming previous reports. The threshold does not depend on the
magnetization direction and therefore the spin Hall injection, which could
indicate an enhancement of the magnon density by the spin Seebeck effect.
However, at negative currents the magnon accumulation remains suppressed which
indicates that the spin Hall effect injection dominates the spin Seebeck
effect. At even higher currents $I\gtrsim I_{\mathrm{DC}}^{\prime\prime}$
another effect kicks in that suppresses the magnon density and conductivity
again. This process is roughly symmetric in the current direction and may be
assigned to a non-linear magnon decay into phonons at elevated temperatures.

Wimmer et al. \cite{1812.01334} also report non-linear effects induced by a Pt
gate current on magnon transport. Their sample is slightly thicker with
13.5\thinspace nm with a damping of $\alpha_{G}=2.17\times10^{-3}$ which is
significantly higher than our $\alpha_{G}=5.2\times10^{-4}$. They report two
anomalies ($I_{\mathrm{on}}$ and $I_{\mathrm{crit}}$). The first appears to
agree with our $I_{\mathrm{DC}}^{\prime}$ and results for $I_{\mathrm{DC}%
}<I_{\mathrm{DC}}^{\prime}$ agree qualitatively with our data and the magnon
conductivity Eq.$\,$(\ref{eq_MagnonSpinConductivity}). The current densities
correspondig to $I_{\mathrm{on}}$ ($3.2\times10^{11}\,$A/m$^{2}$) and
$I_{\mathrm{crit}}$ ($4.3\times10^{11}\,$A/m$^{2}$) are much higher than our
$I_{\mathrm{DC}}^{\prime}$ ($1.1\times10^{11}\,$A/m$^{2}$) and $I_{\mathrm{DC}%
}^{\prime\prime}$ ($2.2\times10^{11}\,$A/m$^{2}$). They do not report
transport for opposite gate current direction and the associated suppression
of the non-local signals, however. For $I_{\mathrm{DC}}>I_{\mathrm{on}}$,
Wimmer et al. \cite{1812.01334} observe signals that increases faster than the
parabolic fit, which we confirm here. However, they do not find the saturation
we report in Figs.$\,$\ref{fig:10nmDevice1V1wMunich}a. Wimmer et al.
\cite{1812.01334} interpret the monotonic increase of their results as an
incipient divergence by an anti-damping spin-orbit torque that compensates the
damping in the YIG film under the gate and speculate about lossless magnon
transport at the onset of self-oscillations or superfluidity. On the other
hand, the larger Gilbert damping in their samples could imply that the magnon
densities at their highest current levels is significantly lower than ours, so
they do not reach the saturation regime that we report here.

Concluding, before drawing conclusion about the nature of nonlinearities, the
complications due to heating should be figured out in more detail
\cite{PhysRevB.97.064422,PhysRevLett.118.127203,optical_mode,PhysRevLett.125.027201}. It would be valuable to assess the
magnon spin accumulation profile governed by the temperature gradient
\cite{PhysRevB.96.184427}, which may be different in thin and thick films. We
conclude that ultrathin YIG films are a great platform for the research on
magnon transport in nonlinear regime, but much work has still to be carried
out before magnon Bose-Einstein condensation or superfluidity by electric or
thermal spin injection can be confirmed.

\section{Acknowledgments}

We acknowledge the helpful discussion with T. Yu and technical support from J.
G. Holstein, H. M. de Roosz, H. Adema T. Schouten and H. de Vries. This work
is part of the research program Magnon Spintronics (MSP) No. 159 financed by
the Foundation for Fundamental Research on Matter (FOM), which is part of the
Netherlands Organisation for Scientific Research (NWO), and supported by the
research programme Skyrmionics with project number 170, which is financed by
the Dutch Research Council (NWO). The support by NanoLab NL is also gratefully
acknowledged. G.B. was supported by JSPS Kakenhi Grant 19H006450.

~\newline

\appendix

\section{Signal change after gate measurement}

\begin{figure}[h]
\includegraphics[width=0.7\linewidth]{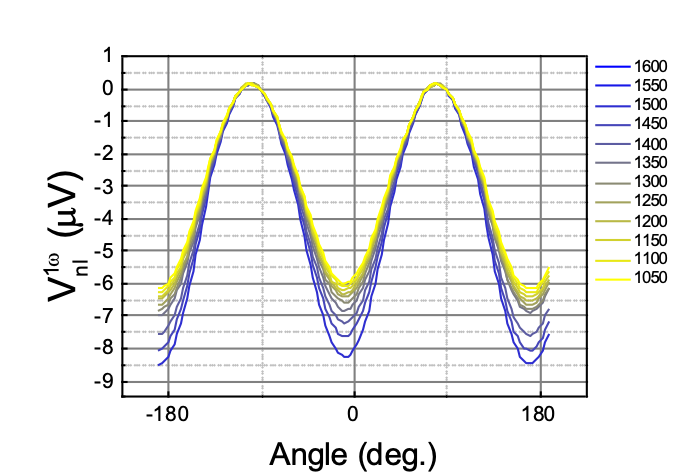}
\caption{Angle-dependent first harmonic voltages at high gate current levels.
The nonlocal signal continues to increase with increasing
current.}%
\label{fig:highdevice2}%
\end{figure}

\begin{figure}[ptbh]
\centering
\begin{subfigure}
		\centering
		\includegraphics[width=0.368\linewidth]{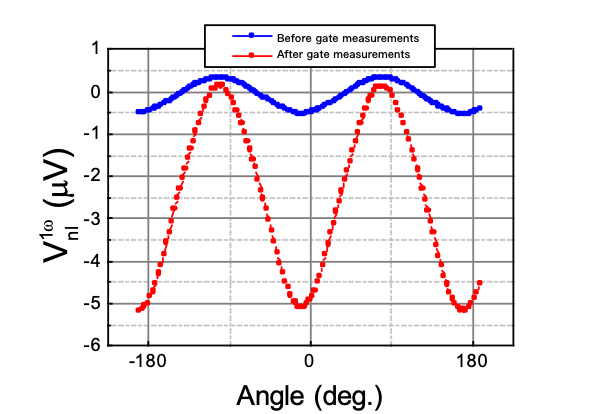}
	\end{subfigure}\hspace*{0.9em} \begin{subfigure}
		\centering
		\includegraphics[width=0.38\linewidth]{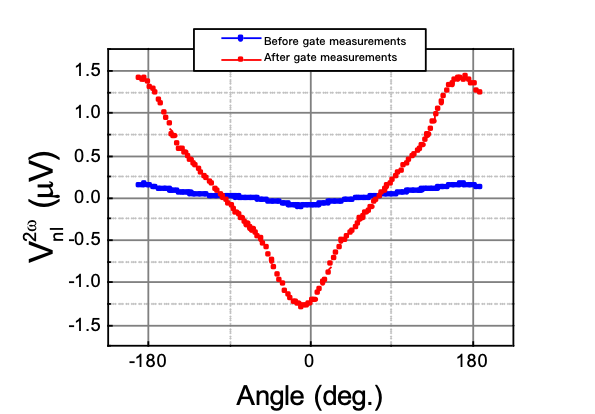}
	\end{subfigure}
\caption{Angle-dependent nonlocal magnon transport measurement before and
after a measurement at high gate currents. The heating that accompanies a
large current changes the properties of YIG. \textbf{a} Angle-dependent first
harmonic voltage before and after the gate-induced heating at zero gate
current. \textbf{b} Angle-dependent second harmonic measurement before and
after. Both first and second harmonic signals are strongly enhanced after the
heating. However, the effect appears to be transient and could not be
reproduced.}%
\label{fig:enhanceddevice2}%
\end{figure}

Device 3 underwent a transient change after applying a high DC current to the
gate. The signal became symmetric around zero angle and enhanced for both 0
and 180 degrees, see Fig.$\,$\ref{fig:highdevice2}, indicating an unidentified
thermal mechanism. After this experiment, the nonlocal signal at zero gate
current increased by a factor five as shown in Fig.$\,$%
\ref{fig:enhanceddevice2}. The high DC current appeared to change the
properties of YIG under the gate. However, after about two weeks, the
characteristics of Device 2 returned back to normal as shown in Fig.$\,$%
\ref{fig:wider_modulator}b.

\section{Modulation effect on 7.9$\,$nm thick YIG}

\begin{figure}[h]
\includegraphics[width=1\linewidth]{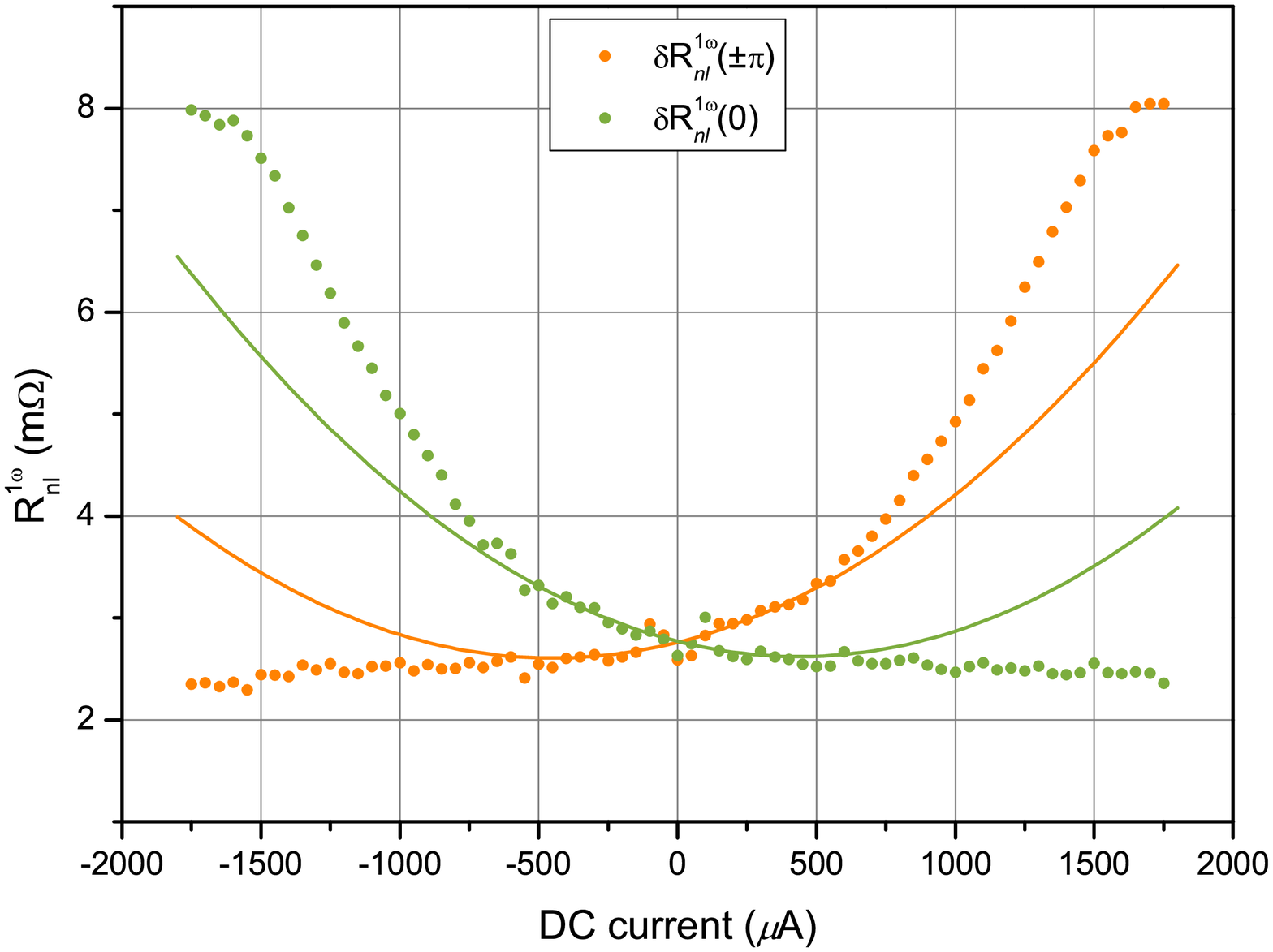} \caption{$R_{\mathrm{nl}%
}^{1\omega}$ for modulator/injector/detector configuration for the 7.9 nm YIG
film specfied in Table \ref{TableDevice6}. Relative amplitudes of the first
harmonic nonlocal signals of device 2 with 800 nm width modulator at
$\alpha=0$ ( \textcolor{orange} {$\bullet$}$R_{\mathrm{nl}}^{1\omega}(0)$) and
$\alpha=\pm\pi$ (\textcolor{ao(english)}{$\bullet$} $R_{\mathrm{nl}}^{1\omega
}(\pm\pi)$) as a function of dc currents.}%
\label{fig:7_9nmyig}%
\end{figure}

We also study a transistor structure on a 7.9$\,$nm thick YIG with damping
parameter of $\alpha_{G}=6.3\times10^{-4}$.  The device parameters are shown
in Table \ref{TableDevice6}. Compared to the 10$\,$nm thick YIG, we observe in
Fig.$\,$\ref{fig:7_9nmyig} a modulation increased by a factor of 3 instead of
2. We have to apply a higher DC currents to reach the nonlinear regime but
still observe a saturation at the highest currents.

\begin{table}[h]
\caption{Geometry of injector/modulator/detector Pt strips.}%
\label{TableDevice6}%
\vspace{3pt} \centering
\setlength{\extrarowheight}{-2pt}
\begin{tabular}
[c]{ccccc}%
\hhline{=====} \vspace{0pt} &  &  &  & \\
Length ($\mu$m) & 20/25/20 &  &  & \\
\vspace{-0.5pt} &  &  &  & \\
Width ($\mu$m) & 0.4/0.4/0.4 &  &  & \\
\vspace{-0.5pt} &  &  &  & \\
Pt thickness (nm) & 8 &  &  & \\
\vspace{-0.5pt} &  &  &  & \\
$I_{\text{ac}}$ ($\mu$A) & 200 &  &  & \\
\vspace{-0.5pt} &  &  &  & \\
$I_{\text{dc}}$ ($m$A) & -1.75$\sim$1.75 &  &  & \\
\vspace{-0.5pt} &  &  &  & \\
Distance between centers of Pt ($\mu$m) & 1.5 &  &  & \\
\vspace{0pt} &  &  &  & \\
\hhline{=====} &  &  &  &
\end{tabular}
\end{table}

\begin{thebibliography}{26}%
\makeatletter
\providecommand \@ifxundefined [1]{%
 \@ifx{#1\undefined}
}%
\providecommand \@ifnum [1]{%
 \ifnum #1\expandafter \@firstoftwo
 \else \expandafter \@secondoftwo
 \fi
}%
\providecommand \@ifx [1]{%
 \ifx #1\expandafter \@firstoftwo
 \else \expandafter \@secondoftwo
 \fi
}%
\providecommand \natexlab [1]{#1}%
\providecommand \enquote  [1]{``#1''}%
\providecommand \bibnamefont  [1]{#1}%
\providecommand \bibfnamefont [1]{#1}%
\providecommand \citenamefont [1]{#1}%
\providecommand \href@noop [0]{\@secondoftwo}%
\providecommand \href [0]{\begingroup \@sanitize@url \@href}%
\providecommand \@href[1]{\@@startlink{#1}\@@href}%
\providecommand \@@href[1]{\endgroup#1\@@endlink}%
\providecommand \@sanitize@url [0]{\catcode `\\12\catcode `\$12\catcode
  `\&12\catcode `\#12\catcode `\^12\catcode `\_12\catcode `\%12\relax}%
\providecommand \@@startlink[1]{}%
\providecommand \@@endlink[0]{}%
\providecommand \url  [0]{\begingroup\@sanitize@url \@url }%
\providecommand \@url [1]{\endgroup\@href {#1}{\urlprefix }}%
\providecommand \urlprefix  [0]{URL }%
\providecommand \Eprint [0]{\href }%
\providecommand \doibase [0]{https://doi.org/}%
\providecommand \selectlanguage [0]{\@gobble}%
\providecommand \bibinfo  [0]{\@secondoftwo}%
\providecommand \bibfield  [0]{\@secondoftwo}%
\providecommand \translation [1]{[#1]}%
\providecommand \BibitemOpen [0]{}%
\providecommand \bibitemStop [0]{}%
\providecommand \bibitemNoStop [0]{.\EOS\space}%
\providecommand \EOS [0]{\spacefactor3000\relax}%
\providecommand \BibitemShut  [1]{\csname bibitem#1\endcsname}%
\let\auto@bib@innerbib\@empty
\bibitem [{\citenamefont {Chumak}\ \emph {et~al.}(2015)\citenamefont {Chumak},
  \citenamefont {Vasyuchka}, \citenamefont {Serga},\ and\ \citenamefont
  {Hillebrands}}]{magnon_spintronics}%
  \BibitemOpen
  \bibfield  {author} {\bibinfo {author} {\bibfnamefont {A.~V.}\ \bibnamefont
  {Chumak}}, \bibinfo {author} {\bibfnamefont {V.~I.}\ \bibnamefont
  {Vasyuchka}}, \bibinfo {author} {\bibfnamefont {A.~A.}\ \bibnamefont
  {Serga}},\ and\ \bibinfo {author} {\bibfnamefont {B.}~\bibnamefont
  {Hillebrands}},\ }\href {https://doi.org/10.1038/nphys3347} {\bibfield
  {journal} {\bibinfo  {journal} {Nature Physics}\ }\textbf {\bibinfo {volume}
  {11}},\ \bibinfo {pages} {453} (\bibinfo {year} {2015})}\BibitemShut
  {NoStop}%
\bibitem [{\citenamefont {Kittel}(1958)}]{ferro}%
  \BibitemOpen
  \bibfield  {author} {\bibinfo {author} {\bibfnamefont {C.}~\bibnamefont
  {Kittel}},\ }\href {https://doi.org/10.1103/PhysRev.110.1295} {\bibfield
  {journal} {\bibinfo  {journal} {Phys. Rev.}\ }\textbf {\bibinfo {volume}
  {110}},\ \bibinfo {pages} {1295} (\bibinfo {year} {1958})}\BibitemShut
  {NoStop}%
\bibitem [{\citenamefont {Eshbach}(1962)}]{ferri}%
  \BibitemOpen
  \bibfield  {author} {\bibinfo {author} {\bibfnamefont {J.~R.}\ \bibnamefont
  {Eshbach}},\ }\href {https://doi.org/10.1103/PhysRevLett.8.357} {\bibfield
  {journal} {\bibinfo  {journal} {Phys. Rev. Lett.}\ }\textbf {\bibinfo
  {volume} {8}},\ \bibinfo {pages} {357} (\bibinfo {year} {1962})}\BibitemShut
  {NoStop}%
\bibitem [{\citenamefont {Lebrun}\ \emph {et~al.}(2018)\citenamefont {Lebrun},
  \citenamefont {Ross}, \citenamefont {Bender}, \citenamefont {Qaiumzadeh},
  \citenamefont {Baldrati}, \citenamefont {Cramer}, \citenamefont {Brataas},
  \citenamefont {Duine},\ and\ \citenamefont {Kl{\"a}ui}}]{antiferro}%
  \BibitemOpen
  \bibfield  {author} {\bibinfo {author} {\bibfnamefont {R.}~\bibnamefont
  {Lebrun}}, \bibinfo {author} {\bibfnamefont {A.}~\bibnamefont {Ross}},
  \bibinfo {author} {\bibfnamefont {S.~A.}\ \bibnamefont {Bender}}, \bibinfo
  {author} {\bibfnamefont {A.}~\bibnamefont {Qaiumzadeh}}, \bibinfo {author}
  {\bibfnamefont {L.}~\bibnamefont {Baldrati}}, \bibinfo {author}
  {\bibfnamefont {J.}~\bibnamefont {Cramer}}, \bibinfo {author} {\bibfnamefont
  {A.}~\bibnamefont {Brataas}}, \bibinfo {author} {\bibfnamefont {R.~A.}\
  \bibnamefont {Duine}},\ and\ \bibinfo {author} {\bibfnamefont
  {M.}~\bibnamefont {Kl{\"a}ui}},\ }\href
  {https://doi.org/10.1038/s41586-018-0490-7} {\bibfield  {journal} {\bibinfo
  {journal} {Nature}\ }\textbf {\bibinfo {volume} {561}},\ \bibinfo {pages}
  {222} (\bibinfo {year} {2018})}\BibitemShut {NoStop}%
\bibitem [{\citenamefont {Oyanagi}\ \emph {et~al.}(2019)\citenamefont
  {Oyanagi}, \citenamefont {Takahashi}, \citenamefont {Cornelissen},
  \citenamefont {Shan}, \citenamefont {Daimon}, \citenamefont {Kikkawa},
  \citenamefont {Bauer}, \citenamefont {van Wees},\ and\ \citenamefont
  {Saitoh}}]{1811.11972}%
  \BibitemOpen
  \bibfield  {author} {\bibinfo {author} {\bibfnamefont {K.}~\bibnamefont
  {Oyanagi}}, \bibinfo {author} {\bibfnamefont {S.}~\bibnamefont {Takahashi}},
  \bibinfo {author} {\bibfnamefont {L.~J.}\ \bibnamefont {Cornelissen}},
  \bibinfo {author} {\bibfnamefont {J.}~\bibnamefont {Shan}}, \bibinfo {author}
  {\bibfnamefont {S.}~\bibnamefont {Daimon}}, \bibinfo {author} {\bibfnamefont
  {T.}~\bibnamefont {Kikkawa}}, \bibinfo {author} {\bibfnamefont {G.~E.~W.}\
  \bibnamefont {Bauer}}, \bibinfo {author} {\bibfnamefont {B.~J.}\ \bibnamefont
  {van Wees}},\ and\ \bibinfo {author} {\bibfnamefont {E.}~\bibnamefont
  {Saitoh}},\ }\href {https://doi.org/10.1038/s41467-019-12749-7} {\bibfield
  {journal} {\bibinfo  {journal} {Nature Communications}\ }\textbf {\bibinfo
  {volume} {10}},\ \bibinfo {pages} {4740} (\bibinfo {year}
  {2019})}\BibitemShut {NoStop}%
\bibitem [{\citenamefont {Uchida}\ \emph {et~al.}(2008)\citenamefont {Uchida},
  \citenamefont {Takahashi}, \citenamefont {Harii}, \citenamefont {Ieda},
  \citenamefont {Koshibae}, \citenamefont {Ando}, \citenamefont {Maekawa},\
  and\ \citenamefont {Saitoh}}]{spin_seebeck}%
  \BibitemOpen
  \bibfield  {author} {\bibinfo {author} {\bibfnamefont {K.}~\bibnamefont
  {Uchida}}, \bibinfo {author} {\bibfnamefont {S.}~\bibnamefont {Takahashi}},
  \bibinfo {author} {\bibfnamefont {K.}~\bibnamefont {Harii}}, \bibinfo
  {author} {\bibfnamefont {J.}~\bibnamefont {Ieda}}, \bibinfo {author}
  {\bibfnamefont {W.}~\bibnamefont {Koshibae}}, \bibinfo {author}
  {\bibfnamefont {K.}~\bibnamefont {Ando}}, \bibinfo {author} {\bibfnamefont
  {S.}~\bibnamefont {Maekawa}},\ and\ \bibinfo {author} {\bibfnamefont
  {E.}~\bibnamefont {Saitoh}},\ }\href {https://doi.org/10.1038/nature07321}
  {\bibfield  {journal} {\bibinfo  {journal} {Nature}\ }\textbf {\bibinfo
  {volume} {455}},\ \bibinfo {pages} {778} (\bibinfo {year}
  {2008})}\BibitemShut {NoStop}%
\bibitem [{\citenamefont {Kajiwara}\ \emph {et~al.}(2010)\citenamefont
  {Kajiwara}, \citenamefont {Harii}, \citenamefont {Takahashi}, \citenamefont
  {Ohe}, \citenamefont {Uchida}, \citenamefont {Mizuguchi}, \citenamefont
  {Umezawa}, \citenamefont {Kawai}, \citenamefont {Ando}, \citenamefont
  {Takanashi}, \citenamefont {Maekawa},\ and\ \citenamefont
  {Saitoh}}]{tunable_spin_hall}%
  \BibitemOpen
  \bibfield  {author} {\bibinfo {author} {\bibfnamefont {Y.}~\bibnamefont
  {Kajiwara}}, \bibinfo {author} {\bibfnamefont {K.}~\bibnamefont {Harii}},
  \bibinfo {author} {\bibfnamefont {S.}~\bibnamefont {Takahashi}}, \bibinfo
  {author} {\bibfnamefont {J.}~\bibnamefont {Ohe}}, \bibinfo {author}
  {\bibfnamefont {K.}~\bibnamefont {Uchida}}, \bibinfo {author} {\bibfnamefont
  {M.}~\bibnamefont {Mizuguchi}}, \bibinfo {author} {\bibfnamefont
  {H.}~\bibnamefont {Umezawa}}, \bibinfo {author} {\bibfnamefont
  {H.}~\bibnamefont {Kawai}}, \bibinfo {author} {\bibfnamefont
  {K.}~\bibnamefont {Ando}}, \bibinfo {author} {\bibfnamefont {K.}~\bibnamefont
  {Takanashi}}, \bibinfo {author} {\bibfnamefont {S.}~\bibnamefont {Maekawa}},\
  and\ \bibinfo {author} {\bibfnamefont {E.}~\bibnamefont {Saitoh}},\ }\href
  {https://doi.org/10.1038/nature08876} {\bibfield  {journal} {\bibinfo
  {journal} {Nature}\ }\textbf {\bibinfo {volume} {464}},\ \bibinfo {pages}
  {262} (\bibinfo {year} {2010})}\BibitemShut {NoStop}%
\bibitem [{\citenamefont {Cornelissen}\ \emph {et~al.}(2015)\citenamefont
  {Cornelissen}, \citenamefont {Liu}, \citenamefont {Duine}, \citenamefont
  {Ben~Youssef},\ and\ \citenamefont {van Wees}}]{Cornelissen2015}%
  \BibitemOpen
  \bibfield  {author} {\bibinfo {author} {\bibfnamefont {L.~J.}\ \bibnamefont
  {Cornelissen}}, \bibinfo {author} {\bibfnamefont {J.}~\bibnamefont {Liu}},
  \bibinfo {author} {\bibfnamefont {R.~A.}\ \bibnamefont {Duine}}, \bibinfo
  {author} {\bibfnamefont {J.}~\bibnamefont {Ben~Youssef}},\ and\ \bibinfo
  {author} {\bibfnamefont {B.~J.}\ \bibnamefont {van Wees}},\ }\href
  {https://doi.org/10.1038/nphys3465} {\bibfield  {journal} {\bibinfo
  {journal} {Nature Physics}\ }\textbf {\bibinfo {volume} {11}},\ \bibinfo
  {pages} {1022} (\bibinfo {year} {2015})}\BibitemShut {NoStop}%
\bibitem [{\citenamefont {Cornelissen}\ \emph {et~al.}(2018)\citenamefont
  {Cornelissen}, \citenamefont {Liu}, \citenamefont {van Wees},\ and\
  \citenamefont {Duine}}]{PhysRevLett.120.097702}%
  \BibitemOpen
  \bibfield  {author} {\bibinfo {author} {\bibfnamefont {L.~J.}\ \bibnamefont
  {Cornelissen}}, \bibinfo {author} {\bibfnamefont {J.}~\bibnamefont {Liu}},
  \bibinfo {author} {\bibfnamefont {B.~J.}\ \bibnamefont {van Wees}},\ and\
  \bibinfo {author} {\bibfnamefont {R.~A.}\ \bibnamefont {Duine}},\ }\href
  {https://doi.org/10.1103/PhysRevLett.120.097702} {\bibfield  {journal}
  {\bibinfo  {journal} {Phys. Rev. Lett.}\ }\textbf {\bibinfo {volume} {120}},\
  \bibinfo {pages} {097702} (\bibinfo {year} {2018})}\BibitemShut {NoStop}%
\bibitem [{\citenamefont {Chumak}\ \emph {et~al.}(2014)\citenamefont {Chumak},
  \citenamefont {Serga},\ and\ \citenamefont
  {Hillebrands}}]{chumak_transistor}%
  \BibitemOpen
  \bibfield  {author} {\bibinfo {author} {\bibfnamefont {A.~V.}\ \bibnamefont
  {Chumak}}, \bibinfo {author} {\bibfnamefont {A.~A.}\ \bibnamefont {Serga}},\
  and\ \bibinfo {author} {\bibfnamefont {B.}~\bibnamefont {Hillebrands}},\
  }\href {https://doi.org/10.1038/ncomms5700} {\bibfield  {journal} {\bibinfo
  {journal} {Nature Communications}\ }\textbf {\bibinfo {volume} {5}},\
  \bibinfo {pages} {4700} (\bibinfo {year} {2014})}\BibitemShut {NoStop}%
\bibitem [{\citenamefont {Wimmer}\ \emph {et~al.}(2019)\citenamefont {Wimmer},
  \citenamefont {Althammer}, \citenamefont {Liensberger}, \citenamefont
  {Vlietstra}, \citenamefont {Gepr\"ags}, \citenamefont {Weiler}, \citenamefont
  {Gross},\ and\ \citenamefont {Huebl}}]{1812.01334}%
  \BibitemOpen
  \bibfield  {author} {\bibinfo {author} {\bibfnamefont {T.}~\bibnamefont
  {Wimmer}}, \bibinfo {author} {\bibfnamefont {M.}~\bibnamefont {Althammer}},
  \bibinfo {author} {\bibfnamefont {L.}~\bibnamefont {Liensberger}}, \bibinfo
  {author} {\bibfnamefont {N.}~\bibnamefont {Vlietstra}}, \bibinfo {author}
  {\bibfnamefont {S.}~\bibnamefont {Gepr\"ags}}, \bibinfo {author}
  {\bibfnamefont {M.}~\bibnamefont {Weiler}}, \bibinfo {author} {\bibfnamefont
  {R.}~\bibnamefont {Gross}},\ and\ \bibinfo {author} {\bibfnamefont
  {H.}~\bibnamefont {Huebl}},\ }\href
  {https://doi.org/10.1103/PhysRevLett.123.257201} {\bibfield  {journal}
  {\bibinfo  {journal} {Phys. Rev. Lett.}\ }\textbf {\bibinfo {volume} {123}},\
  \bibinfo {pages} {257201} (\bibinfo {year} {2019})}\BibitemShut {NoStop}%
\bibitem [{\citenamefont {Shan}\ \emph {et~al.}(2016)\citenamefont {Shan},
  \citenamefont {Cornelissen}, \citenamefont {Vlietstra}, \citenamefont
  {Ben~Youssef}, \citenamefont {Kuschel}, \citenamefont {Duine},\ and\
  \citenamefont {van Wees}}]{PhysRevB.94.174437}%
  \BibitemOpen
  \bibfield  {author} {\bibinfo {author} {\bibfnamefont {J.}~\bibnamefont
  {Shan}}, \bibinfo {author} {\bibfnamefont {L.~J.}\ \bibnamefont
  {Cornelissen}}, \bibinfo {author} {\bibfnamefont {N.}~\bibnamefont
  {Vlietstra}}, \bibinfo {author} {\bibfnamefont {J.}~\bibnamefont
  {Ben~Youssef}}, \bibinfo {author} {\bibfnamefont {T.}~\bibnamefont
  {Kuschel}}, \bibinfo {author} {\bibfnamefont {R.~A.}\ \bibnamefont {Duine}},\
  and\ \bibinfo {author} {\bibfnamefont {B.~J.}\ \bibnamefont {van Wees}},\
  }\href {https://doi.org/10.1103/PhysRevB.94.174437} {\bibfield  {journal}
  {\bibinfo  {journal} {Phys. Rev. B}\ }\textbf {\bibinfo {volume} {94}},\
  \bibinfo {pages} {174437} (\bibinfo {year} {2016})}\BibitemShut {NoStop}%
\bibitem [{\citenamefont {Cornelissen}\ \emph {et~al.}(2016)\citenamefont
  {Cornelissen}, \citenamefont {Peters}, \citenamefont {Bauer}, \citenamefont
  {Duine},\ and\ \citenamefont {van Wees}}]{PhysRevB.94.014412}%
  \BibitemOpen
  \bibfield  {author} {\bibinfo {author} {\bibfnamefont {L.~J.}\ \bibnamefont
  {Cornelissen}}, \bibinfo {author} {\bibfnamefont {K.~J.~H.}\ \bibnamefont
  {Peters}}, \bibinfo {author} {\bibfnamefont {G.~E.~W.}\ \bibnamefont
  {Bauer}}, \bibinfo {author} {\bibfnamefont {R.~A.}\ \bibnamefont {Duine}},\
  and\ \bibinfo {author} {\bibfnamefont {B.~J.}\ \bibnamefont {van Wees}},\
  }\href {https://doi.org/10.1103/PhysRevB.94.014412} {\bibfield  {journal}
  {\bibinfo  {journal} {Phys. Rev. B}\ }\textbf {\bibinfo {volume} {94}},\
  \bibinfo {pages} {014412} (\bibinfo {year} {2016})}\BibitemShut {NoStop}%
\bibitem [{\citenamefont {Ulloa}\ \emph {et~al.}(2019)\citenamefont {Ulloa},
  \citenamefont {Tomadin}, \citenamefont {Shan}, \citenamefont {Polini},
  \citenamefont {van Wees},\ and\ \citenamefont {Duine}}]{1903.02790}%
  \BibitemOpen
  \bibfield  {author} {\bibinfo {author} {\bibfnamefont {C.}~\bibnamefont
  {Ulloa}}, \bibinfo {author} {\bibfnamefont {A.}~\bibnamefont {Tomadin}},
  \bibinfo {author} {\bibfnamefont {J.}~\bibnamefont {Shan}}, \bibinfo {author}
  {\bibfnamefont {M.}~\bibnamefont {Polini}}, \bibinfo {author} {\bibfnamefont
  {B.~J.}\ \bibnamefont {van Wees}},\ and\ \bibinfo {author} {\bibfnamefont
  {R.~A.}\ \bibnamefont {Duine}},\ }\href
  {https://doi.org/10.1103/PhysRevLett.123.117203} {\bibfield  {journal}
  {\bibinfo  {journal} {Phys. Rev. Lett.}\ }\textbf {\bibinfo {volume} {123}},\
  \bibinfo {pages} {117203} (\bibinfo {year} {2019})}\BibitemShut {NoStop}%
\bibitem [{\citenamefont {Chen}\ \emph {et~al.}(2018)\citenamefont {Chen},
  \citenamefont {Roy}, \citenamefont {Cogulu}, \citenamefont {Chang},
  \citenamefont {Wu},\ and\ \citenamefont {Kent}}]{SMR_DC}%
  \BibitemOpen
  \bibfield  {author} {\bibinfo {author} {\bibfnamefont {Y.}~\bibnamefont
  {Chen}}, \bibinfo {author} {\bibfnamefont {D.}~\bibnamefont {Roy}}, \bibinfo
  {author} {\bibfnamefont {E.}~\bibnamefont {Cogulu}}, \bibinfo {author}
  {\bibfnamefont {H.}~\bibnamefont {Chang}}, \bibinfo {author} {\bibfnamefont
  {M.}~\bibnamefont {Wu}},\ and\ \bibinfo {author} {\bibfnamefont {A.~D.}\
  \bibnamefont {Kent}},\ }\href {https://doi.org/10.1063/1.5053120} {\bibfield
  {journal} {\bibinfo  {journal} {Applied Physics Letters}\ }\textbf {\bibinfo
  {volume} {113}},\ \bibinfo {pages} {202403} (\bibinfo {year} {2018})},\
  \Eprint {https://arxiv.org/abs/https://doi.org/10.1063/1.5053120}
  {https://doi.org/10.1063/1.5053120} \BibitemShut {NoStop}%
\bibitem [{\citenamefont {V\'elez}\ \emph {et~al.}(2016)\citenamefont
  {V\'elez}, \citenamefont {Bedoya-Pinto}, \citenamefont {Yan}, \citenamefont
  {Hueso},\ and\ \citenamefont {Casanova}}]{SMR_tem}%
  \BibitemOpen
  \bibfield  {author} {\bibinfo {author} {\bibfnamefont {S.}~\bibnamefont
  {V\'elez}}, \bibinfo {author} {\bibfnamefont {A.}~\bibnamefont
  {Bedoya-Pinto}}, \bibinfo {author} {\bibfnamefont {W.}~\bibnamefont {Yan}},
  \bibinfo {author} {\bibfnamefont {L.~E.}\ \bibnamefont {Hueso}},\ and\
  \bibinfo {author} {\bibfnamefont {F.}~\bibnamefont {Casanova}},\ }\href
  {https://doi.org/10.1103/PhysRevB.94.174405} {\bibfield  {journal} {\bibinfo
  {journal} {Phys. Rev. B}\ }\textbf {\bibinfo {volume} {94}},\ \bibinfo
  {pages} {174405} (\bibinfo {year} {2016})}\BibitemShut {NoStop}%
\bibitem [{\citenamefont {Safranski}\ \emph {et~al.}(2017)\citenamefont
  {Safranski}, \citenamefont {Barsukov}, \citenamefont {Lee}, \citenamefont
  {Schneider}, \citenamefont {Jara}, \citenamefont {Smith}, \citenamefont
  {Chang}, \citenamefont {Lenz}, \citenamefont {Lindner}, \citenamefont
  {Tserkovnyak}, \citenamefont {Wu},\ and\ \citenamefont
  {Krivorotov}}]{spin_auto}%
  \BibitemOpen
  \bibfield  {author} {\bibinfo {author} {\bibfnamefont {C.}~\bibnamefont
  {Safranski}}, \bibinfo {author} {\bibfnamefont {I.}~\bibnamefont {Barsukov}},
  \bibinfo {author} {\bibfnamefont {H.~K.}\ \bibnamefont {Lee}}, \bibinfo
  {author} {\bibfnamefont {T.}~\bibnamefont {Schneider}}, \bibinfo {author}
  {\bibfnamefont {A.~A.}\ \bibnamefont {Jara}}, \bibinfo {author}
  {\bibfnamefont {A.}~\bibnamefont {Smith}}, \bibinfo {author} {\bibfnamefont
  {H.}~\bibnamefont {Chang}}, \bibinfo {author} {\bibfnamefont
  {K.}~\bibnamefont {Lenz}}, \bibinfo {author} {\bibfnamefont {J.}~\bibnamefont
  {Lindner}}, \bibinfo {author} {\bibfnamefont {Y.}~\bibnamefont
  {Tserkovnyak}}, \bibinfo {author} {\bibfnamefont {M.}~\bibnamefont {Wu}},\
  and\ \bibinfo {author} {\bibfnamefont {I.~N.}\ \bibnamefont {Krivorotov}},\
  }\href {https://doi.org/10.1038/s41467-017-00184-5} {\bibfield  {journal}
  {\bibinfo  {journal} {Nature Communications}\ }\textbf {\bibinfo {volume}
  {8}},\ \bibinfo {pages} {117} (\bibinfo {year} {2017})}\BibitemShut {NoStop}%
\bibitem [{\citenamefont {Padr\'on-Hern\'andez}\ \emph
  {et~al.}(2011)\citenamefont {Padr\'on-Hern\'andez}, \citenamefont {Azevedo},\
  and\ \citenamefont {Rezende}}]{Amplification_of_Spin_Waves}%
  \BibitemOpen
  \bibfield  {author} {\bibinfo {author} {\bibfnamefont {E.}~\bibnamefont
  {Padr\'on-Hern\'andez}}, \bibinfo {author} {\bibfnamefont {A.}~\bibnamefont
  {Azevedo}},\ and\ \bibinfo {author} {\bibfnamefont {S.~M.}\ \bibnamefont
  {Rezende}},\ }\href {https://doi.org/10.1103/PhysRevLett.107.197203}
  {\bibfield  {journal} {\bibinfo  {journal} {Phys. Rev. Lett.}\ }\textbf
  {\bibinfo {volume} {107}},\ \bibinfo {pages} {197203} (\bibinfo {year}
  {2011})}\BibitemShut {NoStop}%
\bibitem [{\citenamefont {Bender}\ and\ \citenamefont
  {Tserkovnyak}(2016)}]{Thermally_driven_spin_torques}%
  \BibitemOpen
  \bibfield  {author} {\bibinfo {author} {\bibfnamefont {S.~A.}\ \bibnamefont
  {Bender}}\ and\ \bibinfo {author} {\bibfnamefont {Y.}~\bibnamefont
  {Tserkovnyak}},\ }\href {https://doi.org/10.1103/PhysRevB.93.064418}
  {\bibfield  {journal} {\bibinfo  {journal} {Phys. Rev. B}\ }\textbf {\bibinfo
  {volume} {93}},\ \bibinfo {pages} {064418} (\bibinfo {year}
  {2016})}\BibitemShut {NoStop}%
\bibitem [{\citenamefont {Hamadeh}\ \emph {et~al.}(2014)\citenamefont
  {Hamadeh}, \citenamefont {d'Allivy Kelly}, \citenamefont {Hahn},
  \citenamefont {Meley}, \citenamefont {Bernard}, \citenamefont {Molpeceres},
  \citenamefont {Naletov}, \citenamefont {Viret}, \citenamefont {Anane},
  \citenamefont {Cros}, \citenamefont {Demokritov}, \citenamefont {Prieto},
  \citenamefont {Mu\~noz}, \citenamefont {de~Loubens},\ and\ \citenamefont
  {Klein}}]{PhysRevLett.113.197203}%
  \BibitemOpen
  \bibfield  {author} {\bibinfo {author} {\bibfnamefont {A.}~\bibnamefont
  {Hamadeh}}, \bibinfo {author} {\bibfnamefont {O.}~\bibnamefont {d'Allivy
  Kelly}}, \bibinfo {author} {\bibfnamefont {C.}~\bibnamefont {Hahn}}, \bibinfo
  {author} {\bibfnamefont {H.}~\bibnamefont {Meley}}, \bibinfo {author}
  {\bibfnamefont {R.}~\bibnamefont {Bernard}}, \bibinfo {author} {\bibfnamefont
  {A.~H.}\ \bibnamefont {Molpeceres}}, \bibinfo {author} {\bibfnamefont
  {V.~V.}\ \bibnamefont {Naletov}}, \bibinfo {author} {\bibfnamefont
  {M.}~\bibnamefont {Viret}}, \bibinfo {author} {\bibfnamefont
  {A.}~\bibnamefont {Anane}}, \bibinfo {author} {\bibfnamefont
  {V.}~\bibnamefont {Cros}}, \bibinfo {author} {\bibfnamefont {S.~O.}\
  \bibnamefont {Demokritov}}, \bibinfo {author} {\bibfnamefont {J.~L.}\
  \bibnamefont {Prieto}}, \bibinfo {author} {\bibfnamefont {M.}~\bibnamefont
  {Mu\~noz}}, \bibinfo {author} {\bibfnamefont {G.}~\bibnamefont
  {de~Loubens}},\ and\ \bibinfo {author} {\bibfnamefont {O.}~\bibnamefont
  {Klein}},\ }\href {https://doi.org/10.1103/PhysRevLett.113.197203} {\bibfield
   {journal} {\bibinfo  {journal} {Phys. Rev. Lett.}\ }\textbf {\bibinfo
  {volume} {113}},\ \bibinfo {pages} {197203} (\bibinfo {year}
  {2014})}\BibitemShut {NoStop}%
\bibitem [{\citenamefont {Collet}\ \emph {et~al.}(2016)\citenamefont {Collet},
  \citenamefont {de~Milly}, \citenamefont {d'Allivy Kelly}, \citenamefont
  {Naletov}, \citenamefont {Bernard}, \citenamefont {Bortolotti}, \citenamefont
  {Ben~Youssef}, \citenamefont {Demidov}, \citenamefont {Demokritov},
  \citenamefont {Prieto}, \citenamefont {Mu{\~n}oz}, \citenamefont {Cros},
  \citenamefont {Anane}, \citenamefont {de~Loubens},\ and\ \citenamefont
  {Klein}}]{cite-key}%
  \BibitemOpen
  \bibfield  {author} {\bibinfo {author} {\bibfnamefont {M.}~\bibnamefont
  {Collet}}, \bibinfo {author} {\bibfnamefont {X.}~\bibnamefont {de~Milly}},
  \bibinfo {author} {\bibfnamefont {O.}~\bibnamefont {d'Allivy Kelly}},
  \bibinfo {author} {\bibfnamefont {V.~V.}\ \bibnamefont {Naletov}}, \bibinfo
  {author} {\bibfnamefont {R.}~\bibnamefont {Bernard}}, \bibinfo {author}
  {\bibfnamefont {P.}~\bibnamefont {Bortolotti}}, \bibinfo {author}
  {\bibfnamefont {J.}~\bibnamefont {Ben~Youssef}}, \bibinfo {author}
  {\bibfnamefont {V.~E.}\ \bibnamefont {Demidov}}, \bibinfo {author}
  {\bibfnamefont {S.~O.}\ \bibnamefont {Demokritov}}, \bibinfo {author}
  {\bibfnamefont {J.~L.}\ \bibnamefont {Prieto}}, \bibinfo {author}
  {\bibfnamefont {M.}~\bibnamefont {Mu{\~n}oz}}, \bibinfo {author}
  {\bibfnamefont {V.}~\bibnamefont {Cros}}, \bibinfo {author} {\bibfnamefont
  {A.}~\bibnamefont {Anane}}, \bibinfo {author} {\bibfnamefont
  {G.}~\bibnamefont {de~Loubens}},\ and\ \bibinfo {author} {\bibfnamefont
  {O.}~\bibnamefont {Klein}},\ }\href {https://doi.org/10.1038/ncomms10377}
  {\bibfield  {journal} {\bibinfo  {journal} {Nature Communications}\ }\textbf
  {\bibinfo {volume} {7}},\ \bibinfo {pages} {10377} (\bibinfo {year}
  {2016})}\BibitemShut {NoStop}%
\bibitem [{\citenamefont {Thiery}\ \emph {et~al.}(2018)\citenamefont {Thiery},
  \citenamefont {Naletov}, \citenamefont {Vila}, \citenamefont {Marty},
  \citenamefont {Brenac}, \citenamefont {Jacquot}, \citenamefont {de~Loubens},
  \citenamefont {Viret}, \citenamefont {Anane}, \citenamefont {Cros},
  \citenamefont {Ben~Youssef}, \citenamefont {Beaulieu}, \citenamefont
  {Demidov}, \citenamefont {Divinskiy}, \citenamefont {Demokritov},\ and\
  \citenamefont {Klein}}]{PhysRevB.97.064422}%
  \BibitemOpen
  \bibfield  {author} {\bibinfo {author} {\bibfnamefont {N.}~\bibnamefont
  {Thiery}}, \bibinfo {author} {\bibfnamefont {V.~V.}\ \bibnamefont {Naletov}},
  \bibinfo {author} {\bibfnamefont {L.}~\bibnamefont {Vila}}, \bibinfo {author}
  {\bibfnamefont {A.}~\bibnamefont {Marty}}, \bibinfo {author} {\bibfnamefont
  {A.}~\bibnamefont {Brenac}}, \bibinfo {author} {\bibfnamefont {J.-F.}\
  \bibnamefont {Jacquot}}, \bibinfo {author} {\bibfnamefont {G.}~\bibnamefont
  {de~Loubens}}, \bibinfo {author} {\bibfnamefont {M.}~\bibnamefont {Viret}},
  \bibinfo {author} {\bibfnamefont {A.}~\bibnamefont {Anane}}, \bibinfo
  {author} {\bibfnamefont {V.}~\bibnamefont {Cros}}, \bibinfo {author}
  {\bibfnamefont {J.}~\bibnamefont {Ben~Youssef}}, \bibinfo {author}
  {\bibfnamefont {N.}~\bibnamefont {Beaulieu}}, \bibinfo {author}
  {\bibfnamefont {V.~E.}\ \bibnamefont {Demidov}}, \bibinfo {author}
  {\bibfnamefont {B.}~\bibnamefont {Divinskiy}}, \bibinfo {author}
  {\bibfnamefont {S.~O.}\ \bibnamefont {Demokritov}},\ and\ \bibinfo {author}
  {\bibfnamefont {O.}~\bibnamefont {Klein}},\ }\href
  {https://doi.org/10.1103/PhysRevB.97.064422} {\bibfield  {journal} {\bibinfo
  {journal} {Phys. Rev. B}\ }\textbf {\bibinfo {volume} {97}},\ \bibinfo
  {pages} {064422} (\bibinfo {year} {2018})}\BibitemShut {NoStop}%
\bibitem [{\citenamefont {Qin}\ \emph {et~al.}(2017)\citenamefont {Qin},
  \citenamefont {Zakeri}, \citenamefont {Ernst},\ and\ \citenamefont
  {Kirschner}}]{PhysRevLett.118.127203}%
  \BibitemOpen
  \bibfield  {author} {\bibinfo {author} {\bibfnamefont {H.~J.}\ \bibnamefont
  {Qin}}, \bibinfo {author} {\bibfnamefont {K.}~\bibnamefont {Zakeri}},
  \bibinfo {author} {\bibfnamefont {A.}~\bibnamefont {Ernst}},\ and\ \bibinfo
  {author} {\bibfnamefont {J.}~\bibnamefont {Kirschner}},\ }\href
  {https://doi.org/10.1103/PhysRevLett.118.127203} {\bibfield  {journal}
  {\bibinfo  {journal} {Phys. Rev. Lett.}\ }\textbf {\bibinfo {volume} {118}},\
  \bibinfo {pages} {127203} (\bibinfo {year} {2017})}\BibitemShut {NoStop}%
\bibitem [{\citenamefont {Gepr{\"a}gs}\ \emph {et~al.}(2016)\citenamefont
  {Gepr{\"a}gs}, \citenamefont {Kehlberger}, \citenamefont {Coletta},
  \citenamefont {Qiu}, \citenamefont {Guo}, \citenamefont {Schulz},
  \citenamefont {Mix}, \citenamefont {Meyer}, \citenamefont {Kamra},
  \citenamefont {Althammer}, \citenamefont {Huebl}, \citenamefont {Jakob},
  \citenamefont {Ohnuma}, \citenamefont {Adachi}, \citenamefont {Barker},
  \citenamefont {Maekawa}, \citenamefont {Bauer}, \citenamefont {Saitoh},
  \citenamefont {Gross}, \citenamefont {Goennenwein},\ and\ \citenamefont
  {Kl{\"a}ui}}]{optical_mode}%
  \BibitemOpen
  \bibfield  {author} {\bibinfo {author} {\bibfnamefont {S.}~\bibnamefont
  {Gepr{\"a}gs}}, \bibinfo {author} {\bibfnamefont {A.}~\bibnamefont
  {Kehlberger}}, \bibinfo {author} {\bibfnamefont {F.~D.}\ \bibnamefont
  {Coletta}}, \bibinfo {author} {\bibfnamefont {Z.}~\bibnamefont {Qiu}},
  \bibinfo {author} {\bibfnamefont {E.-J.}\ \bibnamefont {Guo}}, \bibinfo
  {author} {\bibfnamefont {T.}~\bibnamefont {Schulz}}, \bibinfo {author}
  {\bibfnamefont {C.}~\bibnamefont {Mix}}, \bibinfo {author} {\bibfnamefont
  {S.}~\bibnamefont {Meyer}}, \bibinfo {author} {\bibfnamefont
  {A.}~\bibnamefont {Kamra}}, \bibinfo {author} {\bibfnamefont
  {M.}~\bibnamefont {Althammer}}, \bibinfo {author} {\bibfnamefont
  {H.}~\bibnamefont {Huebl}}, \bibinfo {author} {\bibfnamefont
  {G.}~\bibnamefont {Jakob}}, \bibinfo {author} {\bibfnamefont
  {Y.}~\bibnamefont {Ohnuma}}, \bibinfo {author} {\bibfnamefont
  {H.}~\bibnamefont {Adachi}}, \bibinfo {author} {\bibfnamefont
  {J.}~\bibnamefont {Barker}}, \bibinfo {author} {\bibfnamefont
  {S.}~\bibnamefont {Maekawa}}, \bibinfo {author} {\bibfnamefont {G.~E.~W.}\
  \bibnamefont {Bauer}}, \bibinfo {author} {\bibfnamefont {E.}~\bibnamefont
  {Saitoh}}, \bibinfo {author} {\bibfnamefont {R.}~\bibnamefont {Gross}},
  \bibinfo {author} {\bibfnamefont {S.~T.~B.}\ \bibnamefont {Goennenwein}},\
  and\ \bibinfo {author} {\bibfnamefont {M.}~\bibnamefont {Kl{\"a}ui}},\ }\href
  {https://doi.org/10.1038/ncomms10452} {\bibfield  {journal} {\bibinfo
  {journal} {Nature Communications}\ }\textbf {\bibinfo {volume} {7}},\
  \bibinfo {pages} {10452} (\bibinfo {year} {2016})}\BibitemShut {NoStop}%
\bibitem [{\citenamefont {Nambu}\ \emph {et~al.}(2020)\citenamefont {Nambu},
  \citenamefont {Barker}, \citenamefont {Okino}, \citenamefont {Kikkawa},
  \citenamefont {Shiomi}, \citenamefont {Enderle}, \citenamefont {Weber},
  \citenamefont {Winn}, \citenamefont {Graves-Brook}, \citenamefont
  {Tranquada}, \citenamefont {Ziman}, \citenamefont {Fujita}, \citenamefont
  {Bauer}, \citenamefont {Saitoh},\ and\ \citenamefont
  {Kakurai}}]{PhysRevLett.125.027201}%
  \BibitemOpen
  \bibfield  {author} {\bibinfo {author} {\bibfnamefont {Y.}~\bibnamefont
  {Nambu}}, \bibinfo {author} {\bibfnamefont {J.}~\bibnamefont {Barker}},
  \bibinfo {author} {\bibfnamefont {Y.}~\bibnamefont {Okino}}, \bibinfo
  {author} {\bibfnamefont {T.}~\bibnamefont {Kikkawa}}, \bibinfo {author}
  {\bibfnamefont {Y.}~\bibnamefont {Shiomi}}, \bibinfo {author} {\bibfnamefont
  {M.}~\bibnamefont {Enderle}}, \bibinfo {author} {\bibfnamefont
  {T.}~\bibnamefont {Weber}}, \bibinfo {author} {\bibfnamefont
  {B.}~\bibnamefont {Winn}}, \bibinfo {author} {\bibfnamefont {M.}~\bibnamefont
  {Graves-Brook}}, \bibinfo {author} {\bibfnamefont {J.~M.}\ \bibnamefont
  {Tranquada}}, \bibinfo {author} {\bibfnamefont {T.}~\bibnamefont {Ziman}},
  \bibinfo {author} {\bibfnamefont {M.}~\bibnamefont {Fujita}}, \bibinfo
  {author} {\bibfnamefont {G.~E.~W.}\ \bibnamefont {Bauer}}, \bibinfo {author}
  {\bibfnamefont {E.}~\bibnamefont {Saitoh}},\ and\ \bibinfo {author}
  {\bibfnamefont {K.}~\bibnamefont {Kakurai}},\ }\href
  {https://doi.org/10.1103/PhysRevLett.125.027201} {\bibfield  {journal}
  {\bibinfo  {journal} {Phys. Rev. Lett.}\ }\textbf {\bibinfo {volume} {125}},\
  \bibinfo {pages} {027201} (\bibinfo {year} {2020})}\BibitemShut {NoStop}%
\bibitem [{\citenamefont {Shan}\ \emph {et~al.}(2017)\citenamefont {Shan},
  \citenamefont {Cornelissen}, \citenamefont {Liu}, \citenamefont
  {Ben~Youssef}, \citenamefont {Liang},\ and\ \citenamefont {van
  Wees}}]{PhysRevB.96.184427}%
  \BibitemOpen
  \bibfield  {author} {\bibinfo {author} {\bibfnamefont {J.}~\bibnamefont
  {Shan}}, \bibinfo {author} {\bibfnamefont {L.~J.}\ \bibnamefont
  {Cornelissen}}, \bibinfo {author} {\bibfnamefont {J.}~\bibnamefont {Liu}},
  \bibinfo {author} {\bibfnamefont {J.}~\bibnamefont {Ben~Youssef}}, \bibinfo
  {author} {\bibfnamefont {L.}~\bibnamefont {Liang}},\ and\ \bibinfo {author}
  {\bibfnamefont {B.~J.}\ \bibnamefont {van Wees}},\ }\href
  {https://doi.org/10.1103/PhysRevB.96.184427} {\bibfield  {journal} {\bibinfo
  {journal} {Phys. Rev. B}\ }\textbf {\bibinfo {volume} {96}},\ \bibinfo
  {pages} {184427} (\bibinfo {year} {2017})}\BibitemShut {NoStop}%
\end{thebibliography}
\end{document}